\documentclass[a4paper,11pt]{article}
\usepackage{array}
\usepackage{jheppub} 
\usepackage[T1]{fontenc} 
\usepackage{amsmath}   
\usepackage{amssymb}   
\usepackage{booktabs}  
\usepackage{cleveref}
\usepackage{makecell}
\usepackage{subcaption}
\usepackage{physics}

\usepackage[normalem]{ulem}
\usepackage{cellspace} 
\usepackage{slashed}
\title{\boldmath Fusion of Integrable Defects and the Defect $g$-Function}

\author[a,b]{Yang He}
\author[c,d]{Yunfeng Jiang}
\author[a,b]{Yuxiao Liu}

\affiliation[a]{Lanzhou Center for Theoretical Physics, Key Laboratory of Theoretical Physics of Gansu Province,\\
Key Laboratory of Quantum Theory and Applications of MoE,\\
Gansu Provincial Research Center for Basic Disciplines of Quantum Physics,\\
Lanzhou University, Lanzhou 730000, China}
\affiliation[b]{Institute of Theoretical Physics \& Research Center of Gravitation,\\
School of Physical Science and Technology, Lanzhou University, Lanzhou 730000, China}
\affiliation[c]{School of Physics \& Shing-Tung Yau Center, Southeast University, Nanjing 211189, China}
\affiliation[d]{Peng Huanwu Center for Fundamental Theory, Hefei, Anhui 230026, China}

\emailAdd{heyang2023@lzu.edu.cn}
\emailAdd{jiangyf2008@seu.edu.cn}
\emailAdd{liuyx@lzu.edu.cn}

\abstract{We study exact defect $g$-functions for integrable line defects in two-dimensional integrable quantum field theory and use them to probe defect fusion. We consider three settings: fusion of purely transmitting topological defects, fusion of non-topological defects with reflection and transmission, and fusion of a defect with an integrable boundary. For topological defects, the separated logarithmic $g$-function is additive, and the fusion limit is controlled by the multiplicative composition of transmission factors. For non-topological defects, separation-dependent phases in the Bethe-Yang equations produce oscillatory finite-size effects, while the fused defect is described by effective reflection and transmission amplitudes. In the Ising examples studied here, fusion involving non-topological defects lowers the finite localized contribution to the entropy, whereas topological defect-boundary fusion leaves it unchanged.}

\begin{document}
\maketitle
\flushbottom

\section{Introduction}
\label{sec:intro}
Defects are fundamental objects in quantum field theories (QFTs) that describe localized impurities in a bulk system. Beyond modifying bulk physics, as exemplified by phenomena such as the Kondo effect \cite{Kondo:1964nea}, defects encode essential information about the bulk theory. In recent years, topological defects have also emerged as a useful language for describing symmetries, including higher-form \cite{Gaiotto:2014kfa} and non-invertible symmetries \cite{Bhardwaj:2017xup,Chang:2018iay}, offering a unifying framework for traditional and generalized symmetries.\par

In two-dimensional conformal field theories (CFTs), topological line defects have been widely investigated (see \emph{e.g.} \cite{Fuchs:2002cm,Fuchs:2003id,Frohlich:2006ch,Fuchs:2007tx,Buican:2017rxc,Andrei:2018die,Vanhove:2021nav,Northe:2024tnm,Sinha:2026icc,Aasen:2016dop}). By perturbing a CFT with appropriate relevant bulk operators, one can obtain integrable quantum field theories (IQFTs) \cite{Zamolodchikov:1989hfa,Bazhanov:1994ft,Bazhanov:1996aq}. A defect in the original CFT then becomes a defect in the IQFT. However, such a defect usually breaks integrability, unless it is itself perturbed by a relevant defect operator \cite{Runkel:2010ym,Bajnok:2013waa}. When this additional perturbation preserves integrability, the defect is referred to as \emph{integrable}. In the bootstrap approach, an integrable defect is characterized by the transmission and reflection matrices for scattering between bulk particles and the defect, which satisfy the defect Yang-Baxter equations. This description enables the computation of important physical quantities, such as defect partition functions \cite{Martins:1994np,Castro-Alvaredo:2002ulw,Bajnok:2004jd,Bajnok:2007jg,Pozsgay:2010tv}, correlation functions \cite{Bajnok:2009hp,Delfino:1994nx,Delfino:1994nr} in the presence of defects, and entanglement entropy \cite{Jiang:2017qhn,Jiang:2017tyi}.\par

On the other hand, topological line defects can be described by fusion categories \cite{Verlinde:1988sn}, a powerful mathematical framework that plays a central role in the modern description of generalized symmetries \cite{Gaiotto:2014kfa,Chang:2018iay,Schafer-Nameki:2023jdn,McGreevy:2022oyu,Shao:2023gho,Thorngren:2019iar,Bhardwaj:2023kri,Roumpedakis:2022aik,Cordova:2022ruw,Choi:2022jqy,Thorngren:2021yso,Choi:2022zal}. Bridging these two perspectives, namely connecting categorical fusion with the defect Yang-Baxter equations, represents an interesting open problem. Such a connection may also offer a new organizing principle for classifying integrable defects. In a fusion category, a fundamental input is the fusion relation between two defects. To connect the two perspectives, it is useful to understand defect fusion within the bootstrap framework. Related questions have been studied for conformal interfaces and Ising defect lines in \cite{Bachas:2007td,Bachas:2013nxa}, where fusion can require careful subtraction of short-distance singularities.\par

Motivated by this goal, we investigate in this work the fusion of integrable defects in IQFTs. Integrable defects can be broadly classified into two types: topological and non-topological. In interacting theories, integrability imposes strong constraints, forcing scalar integrable defects without internal degrees of freedom to be topological. Non-topological integrable defects may, however, arise in generalized free theories such as free bosons or free fermions. In scattering language, topological defects are purely transmissive and are characterized by a transmission matrix \cite{Konik:1997gx,Bowcock:2003dr}, whereas non-topological defects may involve both transmission and reflection, described by corresponding transmission and reflection matrices. In the limit of vanishing transmission, a defect effectively reduces to a boundary.\par


In this paper, we systematically examine the fusion of defects of both types, with particular attention to the behavior of the defect $g$-function (or defect entropy). This quantity is the defect analogue of the Affleck-Ludwig boundary entropy \cite{Affleck:1991tk}. It measures the non-extensive contribution associated with localized defect degrees of freedom and is therefore a useful probe of defect fusion. More specifically, we investigate the behavior of the defect $g$-function before and after fusion in three situations: (i) fusion of two topological defects; (ii) fusion of two non-topological defects; (iii) fusion of a defect with an integrable boundary.

We find that topological and non-topological defects exhibit qualitatively distinct behaviors under fusion. For topological defects, the $\log g$-function is additive in the separated configuration. The closed- and open-channel descriptions strongly suggest that the finite defect $g$-function is unchanged by fusion, while a complete operator-level defect-TBA proof remains an open problem. In contrast, for non-topological defects, the separation-dependent phase induces strong oscillations and may lead to discontinuities in the $g$-function. The fused defect exhibits smoother behavior and enhanced reflection. In the examples studied below, fusion involving non-topological defects decreases the finite localized contribution to the $g$-function, suggesting a reduction of localized degrees of freedom. When fusing an integrable defect with an integrable boundary, topological defects preserve this finite contribution, whereas non-topological defects lead to a decrease.\par

The structure of the paper is as follows. In Section \ref{sec:ID}, we review integrable defects in IQFT from both conformal perturbation and bootstrap perspectives. We then introduce the defect $g$-function and the thermodynamic Bethe ansatz (TBA), which allows us to compute the defect $g$-function exactly. In Section \ref{sec:one}, we study defect fusion in several cases. We conclude in Section \ref{sec:conclude} with a summary of our results and a discussion of open questions. Technical details of the Bethe-Yang equation derivations are given in Appendix \ref{app:reduce}.

\section{Integrable defects in IQFT}
\label{sec:ID}
In this section, we review two approaches to describing integrable defects in IQFT. The first approach, based on conformal perturbation theory, treats the integrable defect as a suitable perturbation of a topological defect line in a CFT. This description is applicable near the critical point. The second approach is based on an integrable bootstrap perspective, in which the integrable defect is characterized by the scattering matrix between bulk excitations and the defect.

\subsection{Conformal perturbation description}

We consider a renormalization group (RG) flow from a defect conformal field theory (dCFT) to a defect massive IQFT. To produce an integrable defect, the perturbations must be chosen in a manner that preserves an infinite set of conserved charges, even after the bulk conformal symmetry is broken by a mass scale.

The theory is defined on a two-dimensional Euclidean cylinder with coordinates $z = x + iy$ and $\bar{z} = x - iy$, where the spatial direction $x$ is compact: $x\sim x+R$. A defect line is placed along the circle at $y=0$. After performing the conformal mapping $z \to \exp(i z)$ to the complex plane, the defect line is mapped to the unit circle.

We focus on the A-series Virasoro minimal models ${\cal M}(p,p^{\prime})$ where $p,p^{\prime}$ are coprime integers satisfying $p,p^{\prime} \ge 2$. For convenience, we define the ratio $t=p/p^{\prime}$ and $q=\exp(i \pi t)$. The central charge is given by
\begin{equation}
c= 1- \frac{6(p-p^{\prime})^2}{p p^{\prime}}= 1- \frac{6(t-1)^2}{t}\,.
\end{equation}
The operator content is indexed by the set of Kac labels $\cal K$, consisting of integer pairs $(r,s)$ with $1\le r\le p-1$ and $1\le s\le p^{\prime}-1$, subject to the identification $(r,s)\leftrightarrow (p-r,p^{\prime} -s)$. For each independent pair $(r,s)$, there exists an irreducible highest weight representation $R_{(r,s)}$, with the highest weight conformal dimension
\begin{equation}
h_{(r,s)}=\frac{(r-st)^{2}-(1-t)^{2}}{4t} .
\end{equation}
The full Hilbert space decomposes as
\begin{equation}
\mathcal{H}=\bigoplus_{(r,s)\in\mathcal{K}}R_{(r,s)}\otimes \bar{R}_{(r,s)} ,
\end{equation}
where the overline denotes the representation with respect to the anti-holomorphic copy of the Virasoro algebra.

We now consider conformal defects in this setting. Such defects satisfy the constraint \cite{Oshikawa:1996dj}
\begin{equation}
[L_{n}-\bar{L}_{-n},D]=0
\end{equation}
for all integers $n$. In terms of the energy-momentum tensor, this condition physically implies that the energy flow is continuous across the defect. For specific models, such as the Lee-Yang model \cite{Quella:2006de} and the Ising model \cite{Oshikawa:1996dj,Oshikawa:1996ww}, this condition has led to a classification of conformal defects.\par 

\paragraph{Topological defects.} A more stringent constraint requires the independent conservation of the holomorphic and anti-holomorphic components $T(z),\ \bar{T}(\bar{z})$, \emph{i.e.}
\begin{equation}
[L_{n},D]=0=[\bar{L}_{n},D] .
\end{equation}
This defines \emph{topological defects}, also known as \emph{totally transmitting defects} \cite{Bachas:2004sy}. These defects are tensionless and can be freely translated or deformed, provided they do not cross other defects or local operator insertions. 

There exist elementary topological defects $D_{(r,s)}$ labeled by $(r,s) \in \mathcal K$ which cannot be decomposed into a superposition of other defects \cite{Quella:2006de,Runkel:2010ym}. Since these defect operators map each module to itself, by Schur's Lemma they must be proportional to the identity in each irreducible block \cite{Petkova:2000ip}, \emph{i.e.}
\begin{equation}
D_{a}=\sum_{i\in\mathcal{K}}\frac{S_{ai}}{S_{1i}}\cdot{\rm id}_{R_{i}\otimes\bar{R}_{i}} ,
\end{equation}
where $S_{ai}$ denotes the elements of the modular $S$-matrix~\cite{Petkova:2000ip,DiFrancesco:1997nk}, $1$ stands for the Kac labels $(1,1)$ and ${\rm id}_{R_{i}\otimes\bar{R}_{i}}$ is the projector onto the sector $R_{i}\otimes\bar{R}_{i}$. The fusion of two elementary defects $D_{a}$ and $D_{b}$, with $a,b \in \mathcal K$, yields a composite defect $D_{a\star b}\equiv D_{a}D_{b}$, which decomposes according to Verlinde's fusion rules:
\begin{equation}\label{eq:cftfusion}
D_{a\star b} = \sum_{c\in \mathcal K} N_{ab}^{c} \ D_{c} .
\end{equation}
{Here $N_{ab}^{\,c}\in\mathbb{Z}_{\ge 0}$ are the fusion rule coefficients \cite{Verlinde:1988sn}, representing the multiplicity of $D_{c}$ in the decomposition.}\par
Consider a field operator $\phi$ localized on the defect. 
For topological defects $D_{a}$, the separate conservation of the holomorphic and anti-holomorphic energy-momentum tensors ensures that two independent copies of the Virasoro algebra act on the space of defect fields, endowing each with distinct left and right conformal weights $(h,\bar{h})$.

Consequently, the module structure of the defect Hilbert space resembles that of the bulk. However, a key difference is that the spin $h- \bar{h}$ is not constrained to be an integer \cite{Runkel:2010ym}. This arises from the inherent ordering of fields on the one-dimensional defect, which prevents them from being exchanged without passing through one another. As a result, the OPE or commutators exhibit discontinuities when two defect fields coincide, rendering these fields mutually non-local, in contrast to the mutual locality satisfied by the bulk fields.

Moreover, not all the Kac labels $i\in \cal K$ have corresponding defect fields in $D_{a}$. Since defect fields act as intertwiners mapping the defect to itself, a field labeled by $i$ is permitted on the defect $a$ if and only if $i$ appears in the fusion rule of representation $a \star a$.

\paragraph{Non-topological defects.} These defects preserve global energy conservation, but the stricter condition of separate conservation for the holomorphic and anti-holomorphic sectors is relaxed. They can either separate two regions of the same conformal field theory or interface two distinct theories. Even in CFT, such defects have not been fully classified, and the corresponding OPE data is largely unknown. \par
\paragraph{Integrable perturbations.} There are several well-known integrable perturbations for bulk minimal models, such as the primary fields $\Phi_{(1,2)}$, $\Phi_{(1,3)}$. The introduction of a defect slightly complicates the situation.

While integrable perturbations by the primary field $\Phi_{(1,3)}$ can be supported by elementary topological defects, the $\Phi_{(1,2)}$ perturbation is more challenging. Due to the fusion rules of A-type minimal models, the field $\Phi_{(1,2)}$ does not appear in the fusion product of any elementary defect with itself. Consequently, constructing non-local charges for the $\Phi_{(1,2)}$ perturbation requires non-elementary defects, specifically a superposition such as $\mathcal{D} = \mathbf{1} \oplus (1,2)$. This introduces matrix-valued defect fields and significantly more involved consistency conditions \cite{Ambrosino:2025myh} compared with the scalar formalism sufficient for $\Phi_{(1,3)}$. Therefore, we focus on $\Phi_{(1,3)}\equiv \Phi$ to illustrate the construction without these technical complications.
The perturbed Hamiltonian is
\begin{equation}
{H_{P}(\gamma)=L_{0}+\bar{L}_{0}-\frac{c}{12}+\gamma \int_{0}^{2\pi}\Phi(e^{i\theta})\,\dd\theta .}
\end{equation}
{Since the Hamiltonian is defined on the spatial slice $S^1$, the bulk perturbation appears here as an integral over the circle.} The conformal weight of $\Phi$ is $h_{(1,3)}= 2t-1$. To ensure relevance and avoid singularities in OPE \cite{Dotsenko:1984ad,Runkel:1998he}
\begin{equation}
\Phi(z)\Phi(w)~{}=~{}|z-w|^{-4h_{(1,3)}}\,C_{\Phi\Phi}^{~{}{\bf 1}}\cdot{\bf 1}~{}+~{}|z-w|^{-2h_{(1,3)}}\,C_{\Phi\Phi}^{~{}\Phi}\cdot\Phi(w)~{}+~{}\dots ,
\end{equation}
the ratio $t$ is chosen such that $t<1/2$ in the following discussion, which makes the theory non-unitary. The field $\Phi$ is normalized such that the structure constants are
\begin{equation}
\begin{split}
C_{\Phi\Phi}^{~{}{\bf 1}} &=\frac{\sin(3\pi t)}{\sin(\pi t)}\cdot\eta_{\Phi}^{\,2}\ ,\\
C_{\Phi\Phi}^{~{}\Phi} &=-\frac{(1{-}2t)(1{-}3t)}{1{-}4t}\,\frac{\Gamma\big{(}t\big{)}^{2}\,\Gamma\big{(}1{-}2t\big{)}^{2}\,\Gamma\big{(}4t\big{)}}{\Gamma\big{(}1{-}t\big{)}\,\Gamma\big{(}2t\big{)}^{2}\,\Gamma\big{(}3t\big{)}\,\Gamma\big{(}2{-}4t\big{)}}\cdot\eta_{\Phi}\ .
\end{split}
\end{equation}

In general, the pure bulk perturbation of the dCFT by $\Phi$ does not render the defect integrable, as momentum conservation is violated. To have integrable defects, one must consider specific defect perturbations. For a given integrable bulk perturbation, determining which defect perturbations preserve integrability is non-trivial. 
For non-topological defects, it is convenient to fold the system into a boundary CFT (BCFT) and perturb the resulting boundary. While a complete classification is open for general CFTs, the problem is well understood for specific cases such as the free boson \cite{Bachas:2001vj} and the Ising model \cite{Oshikawa:1996dj,Oshikawa:1996ww,Kormos:2009sk}, where continuous families of non-topological conformal defects were identified using boundary states in the folded theory. Reflection, transmission and entropy for conformal perturbation defects have also been studied in \cite{Konechny:2014fva,Brunner:2015vva}. \par

For topological defects, this question has been investigated for the model under consideration in \cite{Bajnok:2013waa,Runkel:2010ym,Ambrosino:2025myh,Runkel:2007wd,Manolopoulos:2009np}. 
Because a topological defect preserves two copies of the Virasoro algebra, defect fields can be organized into Virasoro representations, which allows one to classify candidate defect operators and perturbations. Moreover, to maintain integrability one typically chooses the defect perturbing fields to be the same relevant operator that generates the integrable bulk deformation.

Henceforth, we abbreviate $(1,s)$ as $s$. For ${\cal M}(p,p^{\prime})$ with $t<1/2$ and the bulk perturbation $\Phi$, we focus on the defect $D_{s}$ and the defect primary field $\phi^{s}$ and $\bar{\phi}^{s}$ with weights $(h_{(1,3)},0)$ and $(0, h_{(1,3)})$ respectively. The OPEs for these fields are \cite{Fuchs:2004xi}
\begin{equation}
\begin{split}
\phi^{s}(x)\phi^{s}(y)~{}=~{}(x{-}y)^{-2h_{(1,3)}}\,C^{(s){\bf 1}}_{\phi\phi}\cdot{\bf 1}^{s}~{}+~{}(x{-}y)^{-h_{(1,3)}}\,C^{(s)\phi}_{\phi\phi}\cdot\phi^{s}(y)~{}+\,\dots\ ,\\
\bar{\phi}^{s}(x)\bar{\phi}^{s}(y)~{}=~{}(x{-}y)^{-2h_{(1,3)}}\,C^{(s){\bf 1}}_{\phi\phi}\cdot{\bf 1}^{s}~{}+~{}(x{-}y)^{-h_{(1,3)}}\,C^{(s)\phi}_{\phi\phi}\cdot\bar{\phi}^{s}(y)~{}+\,\dots\ ,
\end{split}
\end{equation}
where ${\bf 1}^{s}$ is the identity field on the $(1,s)$ defect line. The structure constants are given by
\begin{equation}
\begin{split}
C^{(s){\bf 1}}_{\phi\phi} &=-\frac{1-(s{+}1)t}{1{-}3t}\,\frac{\Gamma\big{(}2-(1{+}s)t\big{)}\,\Gamma\big{(}(s{-}1)t\big{)}\,\Gamma\big{(}1{-}t\big{)}\,\Gamma\big{(}2t\big{)}\,\Gamma\big{(}3t\big{)}}{\Gamma\big{(}(1{+}s)t\big{)}\,\Gamma\big{(}1-(s{-}1)t\big{)}\,\Gamma\big{(}2{-}2t\big{)}\,\Gamma\big{(}t\big{)}^{2}}\cdot\big{(}\eta_{\phi}^{(s)}\big{)}^{2}\ ,\\
C^{(s)\phi}_{\phi\phi} &=\frac{2\sin(\pi t)\cos(\pi st)}{\sin(4\pi t)}\,\frac{\Gamma\big{(}2-(s{+}1)t\big{)}\,\Gamma\big{(}(s{-}1)t\big{)}}{\Gamma\big{(}2{-}4t\big{)}\,\Gamma\big{(}2t\big{)}}\cdot\eta_{\phi}^{(s)}\ .
\end{split}
\end{equation}

We consider the relevant defect field $\psi_{\lambda,\tilde{\lambda}}^{s}=\lambda \phi^{s} + \tilde{\lambda} \bar{\phi}^{s}$, which perturbs $D_{s}$ to
\begin{equation}
D_{s}(\psi_{\lambda,\tilde{\lambda}}^{s}) = \sum_{n=0}^{\infty} \frac{1}{n!} D_{s}^{(n)}(\psi_{\lambda,\tilde{\lambda}}^{s})\ ,
\end{equation}
where $\lambda$ and $\tilde{\lambda}$ are complex parameters. The $n$-th order term $D_{s}^{(n)}(\psi)$ is defined as
\begin{equation}
D_{s}^{(n)}(\psi)=\int_{0}^{2\pi} D_{s}[\psi(\theta_{1}),\dots,\psi(\theta_{n})]\,\dd\theta_{1}\cdots \dd\theta_{n}\ .
\end{equation}
Here, $D_{s}[\psi(\theta_{1}),\dots,\psi(\theta_{n})]$ represents a specific configuration where the defect forms a unit circle with clockwise orientation and defect fields $\psi_{k}$ are inserted at $\exp(i\theta_{k})$. 

\paragraph{Fusion of defects.} For these perturbed defects, fusion rules persist but generally do not decompose into a direct sum. For a fused defect $D_{a\star b} (\psi{\times}{\bf 1}^{b}+{\bf 1}^{a}{\times}\phi) \equiv D_{a}(\psi)D_{b}(\phi)$, if $a\star b = \bigoplus_{i} c_{i}$ with $N_{ab}^{c_{i}}=1$ for all $c_{i}$, the decomposition \cite{Runkel:2007wd,Manolopoulos:2009np} can be written as
\begin{equation}
D_{a\star b}(\psi{\times}{\bf 1}^{b}+{\bf 1}^{a}{\times}\phi)=D_{\oplus_i c_i}({\textstyle\sum_{i,j}}\,\xi^{c_{i}\rightarrow c_{j}})\ .
\end{equation}
In this context, $\xi^{c_{i}\rightarrow c_{j}}$ represents the projection of the total perturbation onto the transition channel between the defect components $c_i$ and $c_j$. Consequently, a non-vanishing $\xi^{c_{i}\rightarrow c_{j}}$ for $i \neq j$ implies that the perturbed defect operator cannot be written as a direct sum of independent operators, as the perturbation mixes the different defect sectors.

Consider the composition of the defect operator $D_2$ with a generic defect $D_s$. In general, this fusion excites all permissible channels in the decomposition $2 \star s = (s-1) \oplus (s+1)$
\begin{equation}
\begin{split}
&D_{2}(\psi_{\lambda,\tilde{\lambda}})D_{s}(\psi_{\mu,\tilde{\mu}})\\
=&D_{2\star s}(\lambda\phi^{2}{\times}{\bf 1}^{s}+\mu{\bf 1}^{2}{\times}\phi^{s}+\tilde{\lambda}\bar{\phi}^{2}{\times}{\bf 1}^{s}+\tilde{\mu}{\bf 1}^{2}{\times}\bar{\phi}^{s})\\
=&D_{s{-}1\oplus s{+}1}({\sum_{\sigma,\nu=\pm 1}}~{}\xi_{\sigma,\nu}\cdot\phi^{s+\nu\rightarrow s+\sigma}+\bar{\xi}_{\sigma,\nu}\cdot\bar{\phi}^{s+\nu\rightarrow s+\sigma})\ .
\end{split}
\end{equation}
Here, the fields $\phi^{s+\nu\rightarrow s+\sigma}$ serve as transition operators mixing the defect sectors, analogous to $\xi^{c_i \to c_j}$ discussed above.

By setting $\mu = q^{s \varepsilon} \lambda$, {$\tilde{\mu} = q^{-s \varepsilon} \tilde{\lambda}$} ($\varepsilon = \pm 1$) and choosing the appropriate normalization of CFT structure constants 
\begin{equation}
\eta_{\phi}^{(s)}=\frac{\Gamma\big{(}t\big{)}\Gamma\big{(}2{-}3t\big{)}}{\Gamma\big{(}(s{-}1)t\big{)}\Gamma\big{(}2{-}(s{+}1)t\big{)}}\cdot\eta_{\phi}^{(2)}\ ,
\end{equation}
the coefficients of the off-diagonal transition fields vanish \cite{Runkel:2007wd,Manolopoulos:2009np}. Consequently, the perturbed defect operator decomposes into a sum of independent operators on $D_{s-1}$ and $D_{s+1}$
\begin{equation}
\begin{split}
&D_{2}(\psi_{\lambda,\tilde{\lambda}})\,D_{s}(\psi_{q^{s\varepsilon}\lambda,q^{-s\varepsilon}\tilde{\lambda}})\\
&\quad = D_{s-1}(\psi_{q^{(s+1)\varepsilon}\lambda,q^{-(s+1)\varepsilon}\tilde{\lambda}})~{}+~{}D_{s+1}(\psi_{q^{(s-1)\varepsilon}\lambda,q^{-(s-1)\varepsilon}\tilde{\lambda}})\ .
\end{split}
\end{equation}
This recurrence relation reveals that the fundamental defect $D_2$ acts as a generating operator for all higher-spin defect operators \cite{Runkel:2010ym}. Furthermore, this identity implies the T-system functional relation recursively,
\begin{equation}\label{eq:same_defect_fusion}
D_{s}(\psi_{q\lambda,q^{-1}\tilde{\lambda}})\,D_{s}(\psi_{q^{-1}\lambda,q\tilde{\lambda}})~{}=~{}{\rm id}+D_{s-1}(\psi_{\lambda,\tilde{\lambda}})D_{s+1}(\psi_{\lambda,\tilde{\lambda}})\ .
\end{equation}
This relation governs the fusion of identical perturbed defects at shifted spectral parameters {$q\lambda , q^{-1}\tilde{\lambda}$ and $q^{-1}\lambda , q \tilde{\lambda}$ respectively}, a characteristic feature of the integrable structure inherent in the perturbed conformal field theory.

\paragraph{Integrable defect perturbation.} 
{Following the approach based on non-local conserved charges \cite{Bazhanov:1994ft,Bazhanov:1996aq}, an integrable bulk perturbation can be studied via QFT transfer matrices, which generate an infinite family of mutually commuting conserved charges. In the defect framework \cite{Runkel:2010ym,Runkel:2007wd,Manolopoulos:2009np}, these transfer matrices can be interpreted as defect operators, which inherit the hallmark properties of transfer matrices, such as mutual commutativity and the T-system fusion relations. Accordingly, integrability of a defect perturbation means that the perturbed defect operator continues to generate an infinite set of commuting conserved charges. We summarize three key properties below.} First, the defect operators mutually commute
\begin{equation}
\big{[}\,D_{m}(\psi_{\lambda,\tilde{\lambda}})\,,\,D_{n}(\psi_{\mu,\tilde{\mu}})\,\big{]}~{}=~{}0\ ,
\end{equation}
provided $\lambda \tilde{\lambda} = \mu \tilde{\mu}$. Second, when
\begin{equation}
\xi=\frac{\big{(}\eta^{(2)}_{\phi}\big{)}^{2}}{\eta_{\Phi}}\,\frac{-1}{\sin(3\pi t)}\,\frac{\Gamma\big{(}2{-}3t\big{)}\Gamma\big{(}1{-}t\big{)}}{\Gamma\big{(}2{-}2t\big{)}\Gamma\big{(}1{-}2t\big{)}}\ ,
\end{equation}
the Hamiltonian satisfies
\begin{equation}
\big{[}\,{H_{P}(\xi \lambda \tilde{\lambda})} \,,\,D_{s}(\psi_{\lambda,\tilde{\lambda}})\,\big{]}~{}=~{}0\ .
\end{equation}
The third property facilitates the identification of UV and IR parameters using BCFT. The physical intuition is that defect fields become boundary fields when the defect approaches the boundary. 

For an elementary defect field, the holomorphic and anti-holomorphic parts merge into the same boundary field $\lambda \phi^{s}+\tilde{\lambda} \bar{\phi}^{s} \to (\lambda+\tilde{\lambda})\chi^{s}$. Let $|(1,1)\rangle\!\rangle$ denote the Cardy boundary state of the vacuum representation, and $|(1,s)+\gamma \chi^s\rangle\!\rangle$ denote the state corresponding to the $(1,s)$ boundary condition perturbed by $\gamma \chi^{s}$ on the complex plane with the unit disk removed. Thus, we have 
\begin{equation}
D_{s}(\psi_{\lambda,\tilde{\lambda}})|(1,1)\rangle\!\rangle = |(1,s)+ (\lambda+\tilde{\lambda}) \chi^s\rangle\!\rangle.
\end{equation}

\subsection{Bootstrap description}

We now turn to a description more commonly used in 1+1-dimensional IQFTs. We adopt the scattering formalism, in which the dynamics is characterized by the factorized S-matrix of the particles, along with the scattering between particles and the defect. A useful tool in this context is the Zamolodchikov-Faddeev (ZF) operator, which satisfies an algebra \cite{Zamolodchikov:1978xm} that generalizes the standard Fock space commutation relations by incorporating the exact $S$-matrix.

Let $A_i(\theta)$ denote the ZF creation operator for a particle of species $i$ with rapidity $\theta$. It satisfies the fundamental commutation relations
\begin{align}
A_i(\theta_1) A_j(\theta_2) &= S_{ij}^{kl}(\theta_{12}) A_k(\theta_2) A_l(\theta_1), \label{eq:bulk_comm}
\end{align}
where $\theta_{12} \equiv \theta_1 - \theta_2$. Applying this commutation relation twice yields unitarity condition for the bulk $S$-matrix
\begin{equation}
S_{ij}^{kl}(\theta)S_{kl}^{mn}(-\theta) = \delta_{i}^{m}\delta_{j}^{n} .
\end{equation}

We introduce a defect operator $D_\alpha$ \cite{Delfino:1994nx,Delfino:1994nr,Castro-Alvaredo:2002qcm} to describe the line defect. The index $\alpha$ characterizes the internal state of the defect. Note that this operator is distinct from the defect operator discussed in the previous section. The interaction between bulk excitations and the defect is encoded in an extended algebra involving $A_i(\theta)$ and $D_{\alpha}$, given by
\begin{align}
A_i(\theta) D_\alpha &= R_{i\alpha}^{j\beta}(\theta) A_j(-\theta) D_\beta + T_{i\alpha}^{j\beta}(\theta) D_\beta A_j(\theta), \label{eq:defect_left}\\
D_\alpha A_i(-\theta)  &= \tilde{R}_{i\alpha}^{j\beta}(\theta) D_\beta A_j(\theta)  + \tilde{T}_{i\alpha}^{j\beta}(\theta) A_j(-\theta) D_\beta. \label{eq:defect_right}
\end{align}
Here, $A_i(\theta)$ and $A_i(-\theta)$ describe right-moving and left-moving particles, respectively. The operators $R$ and $T$ denote the reflection and transmission amplitudes for particles incident from the left, while $\tilde{R}$ and $\tilde{T}$ correspond to incidence from the right.

Applying this algebra twice and invoking the linear independence of the states (e.g., $A_i(\theta) D_\alpha$, $A_k(\theta) D_\gamma$, and $D_\gamma A_k(-\theta)$), we derive the unitarity constraints for the reflection and transmission amplitudes:
\begin{equation}
\begin{split}
R_{i \alpha}^{j \beta}(\theta) R_{j \beta}^{k \gamma}(-\theta)+T_{i \alpha}^{j \beta}(\theta) \tilde{T}_{j \beta}^{k \gamma}(-\theta) & =\delta_{i}^{k} \delta_{\alpha}^{\gamma}, \\
R_{i \alpha}^{j \beta}(\theta) T_{j \beta}^{k \gamma}(-\theta)+T_{i \alpha}^{j \beta}(\theta) \tilde{R}_{j \beta}^{k \gamma}(-\theta) & =0\,.
\end{split}
\end{equation}

\begin{figure}[tbp]
    \centering
    \begin{subfigure}[b]{0.48\linewidth}
        \centering
        \includegraphics[width=0.65\textwidth]{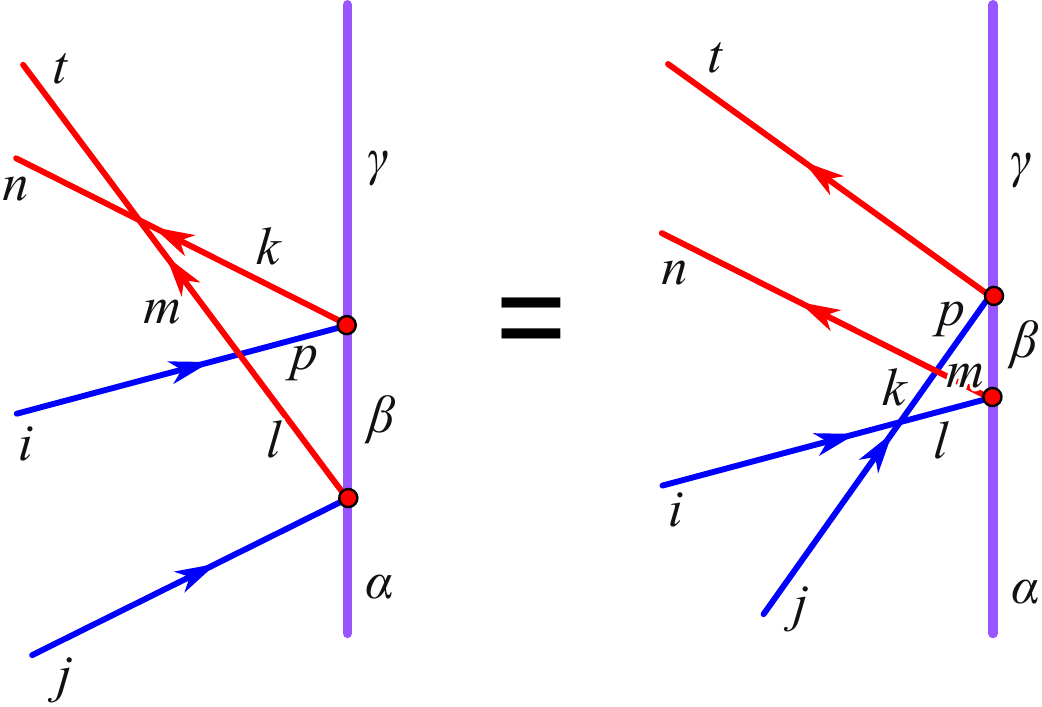}
        \caption{Pure reflection}
        \label{fig:reflection}
    \end{subfigure}
    \hfill
    \begin{subfigure}[b]{0.48\linewidth}
        \centering
        \includegraphics[width=1\textwidth]{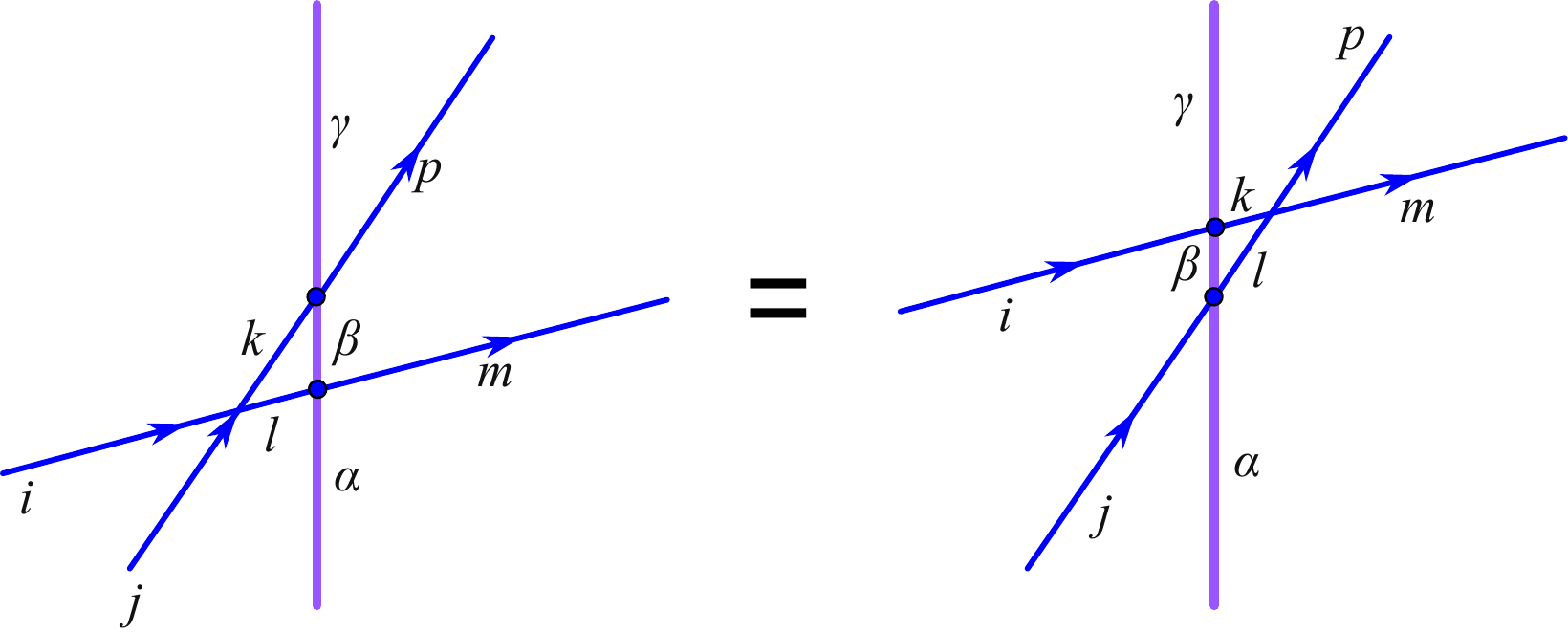}
        \caption{Pure transmission}
        \label{fig:transmission}
    \end{subfigure}
    
    \vspace{0.8em}

    \begin{subfigure}[b]{0.48\linewidth}
        \centering
        \includegraphics[width=\textwidth]{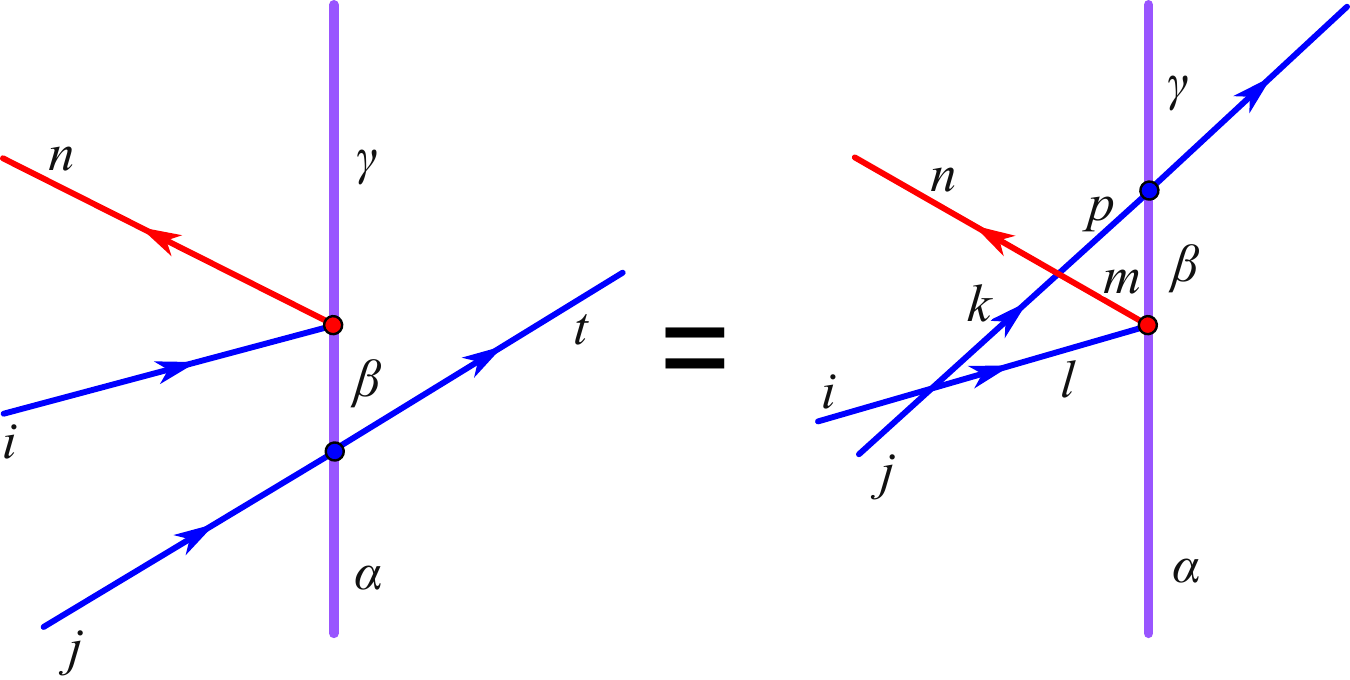}
        \caption{Mixed scattering 1}
        \label{fig:mixed1}
    \end{subfigure}
    \hfill 
    \begin{subfigure}[b]{0.48\linewidth}
        \centering
        \includegraphics[width=\textwidth]{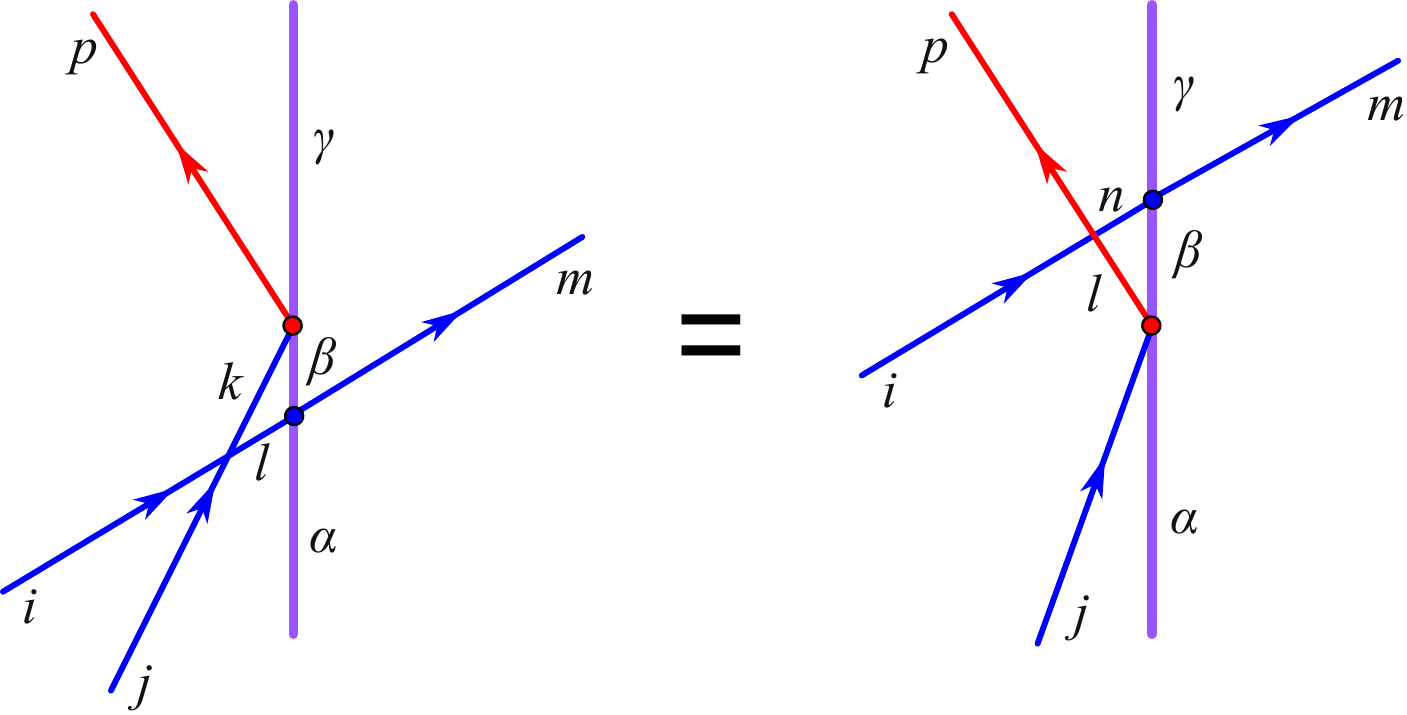}
        \caption{Mixed scattering 2}
        \label{fig:mixed2}
    \end{subfigure}    
    \caption{{Schematic illustration of elementary defect-scattering processes in the presence of a localized defect. The vertical magenta line denotes the defect worldline, while the oriented external legs represent asymptotic particle trajectories. Blue lines correspond to incoming or transmitted particles, whereas red lines indicate reflected particles. Panel~(a) shows pure reflection, in which the particle is bounced back by the defect and remains on the same side. Panel~(b) shows pure transmission, in which the particle crosses the defect and continues to the opposite side without reversing its direction of motion. Panels~(c) and~(d) illustrate mixed processes in which both transmission and reflection are present as allowed scattering channels; the two diagrams distinguish the two possible orderings of these elementary defect interactions, namely transmission followed by reflection and reflection followed by transmission.}}
    \label{fig:defect_scattering_processes}
\end{figure}

\paragraph{Defect Yang-Baxter equations.} The associativity of the ZF algebra imposes the defect Yang-Baxter equations, ensuring that the outcome of three-body scattering processes is independent of the order of interaction. For instance, consider the initial state $A_{i}(\theta_{1})A_{j}(\theta_{2})D_{\alpha}$. The purely reflected contribution, corresponding to Fig.~\ref{fig:defect_scattering_processes}\subref{fig:reflection}, is obtained from the coefficient of $A_{i}(-\theta_{1})A_{j}(-\theta_{2})D_{\alpha}$ and yields the constraint
\begin{equation}
R_{j\alpha}^{l\beta}(\theta_{2})S_{il}^{mp}(\hat{\theta}_{12})R_{p\beta}^{k\gamma}(\theta_{1})S_{mk}^{nt}(\theta_{12}) = S_{ij}^{kl}(\theta_{12})R_{l\alpha}^{m\beta}(\theta_{1})S_{km}^{np}(\hat{\theta}_{12})R_{p\beta}^{t\gamma}(\theta_{2})\ ,
\end{equation}
where we denote the sum of rapidities as $\hat{\theta}_{12}\equiv\theta_{1}+\theta_{2}$. Similarly, the purely transmissive contribution, corresponding to Fig.~\ref{fig:defect_scattering_processes}\subref{fig:transmission}, is extracted from the coefficient of $D_{\alpha}A_{j}(\theta_{2})A_{i}(\theta_{1})$, leading to
\begin{equation}
S_{ij}^{kl}(\theta_{12})T_{l\alpha}^{m\beta}(\theta_{1})T_{k\beta}^{p\gamma}(\theta_{2}) = T_{j\alpha}^{l\beta}(\theta_{2})T_{i\beta}^{k\gamma}(\theta_{1})S_{kl}^{pm}(\theta_{12})\ .
\end{equation}
The mixed terms, corresponding to the two orderings shown in Fig.~\ref{fig:defect_scattering_processes}\subref{fig:mixed1} and Fig.~\ref{fig:defect_scattering_processes}\subref{fig:mixed2}, arise from the coefficients of $A_{i}(-\theta_{1})D_{\alpha}A_{j}(\theta_{2})$ and $A_{i}(-\theta_{2})D_{\alpha}A_{j}(\theta_{1})$, respectively, and are given by
\begin{equation}
\begin{split}
T_{j\alpha}^{t\beta}(\theta_{2})R_{i\beta}^{n\gamma}(\theta_{1}) &= S_{ij}^{kl}(\theta_{12})R_{l\alpha}^{m\beta}(\theta_{1})S_{km}^{np}(\hat{\theta}_{12})T_{p\beta}^{t\gamma}(\theta_{2}) \ ,\\
S_{ij}^{kl}(\theta_{12})T_{l\alpha}^{m\beta}(\theta_{1})R_{k\beta}^{p\gamma}(\theta_{2}) &= R_{j\alpha}^{l\beta}(\theta_{2})S_{il}^{pn}(\hat{\theta}_{12})T_{n\beta}^{m\gamma}(\theta_{1})\ .
\end{split}
\end{equation}
Note that these two equations are related by applying $S(\theta_{21})$ and exchanging the rapidities.

Different constraints arise from initial states with opposite parity, like $D_{\alpha} A_{i}(\theta_1) A_{j}(\theta_2)$, or from separated states like $A_{i}(\theta_{1})D_{\alpha}A_{j}(\theta_{2})$. The former yields results identical to those above under the replacements $T \to \tilde{T}$ and $R \to \tilde{R}$. The latter leads to a distinct set of constraints obtained via the same method:
\begin{align}
R_{i\alpha}^{k\beta}(\theta_{1})\tilde{R}_{j\beta}^{l\gamma}(\theta_{2}) &= R_{i\beta}^{k\gamma}(\theta_{1})\tilde{R}_{j\alpha}^{l\beta}(\theta_{2}), \\
\tilde{T}_{l\beta}^{q\gamma}(\theta_{2})S_{kj}^{lp}(\hat{\theta}_{12})T_{i\alpha}^{k\beta}(\theta_{1}) &=   T_{t\beta}^{p\gamma}(\theta_{1})S_{ik}^{qt}(\hat{\theta}_{12})\tilde{T}_{j\alpha}^{k\beta}(\theta_{2}),\\
\tilde{R}_{l\beta}^{q\gamma}(\theta_{2})S_{kj}^{lp}(\hat{\theta}_{12})T_{i\alpha}^{k\beta}(\theta_{1}) &= S_{lk}^{qp}(\theta_{12})T_{i\beta}^{l\gamma}(\theta_{1})\tilde{R}_{j\alpha}^{k\beta}(\theta_{2}), \\
S_{lq}^{st}(\theta_{12})R_{p\beta}^{q\gamma}(\theta_{1})S_{ik}^{lp}(\hat{\theta}_{12})\tilde{T}_{j\alpha}^{k\beta}(\theta_{2}) &=   R_{i\alpha}^{s\beta}(\theta_{1})\tilde{T}_{j\beta}^{t\gamma}(\theta_{2})\ ,
\end{align}
corresponding to the final states $A_{i}(-\theta_{1})D_{\alpha}A_{j}(-\theta_{2})$, $A_{j}(\theta_{2})D_{\alpha}A_{i}(\theta_{1})$, $D_{\alpha}A_{j}(-\theta_{2})A_{i}(\theta_{1})$ and $A_{i}(-\theta_{1})A_{j}(\theta_{2})D_{\alpha}$ respectively.

\paragraph{A no-go theorem.} Collectively, these relations constrain the form of reflection and transmission amplitudes. In particular, they lead to the following no-go theorem: \emph{if the defect carries no internal degrees of freedom, then the underlying theory must be a generalized free field theory characterized by a rapidity-independent $S$-matrix to allow for both reflection and transmission} \cite{Delfino:1994nx,Delfino:1994nr,Castro-Alvaredo:2002qcm}. Conversely, if the defect possesses internal degrees of freedom, simultaneous reflection and transmission can occur only in non-diagonal bulk theories, provided that the following exchange relation is violated \cite{Castro-Alvaredo:2002qcm}
\begin{equation}
T_{\alpha}^{\beta}(\theta_{1}) \otimes R_{\beta}^{\gamma}(\theta_{2}) = T_{\beta}^{\gamma}(\theta_{1}) \otimes R_{\alpha}^{\beta}(\theta_{2}).
\end{equation}

In the limit of vanishing transmission, the defect operator is replaced by a boundary operator $B_\alpha$ \cite{Cherednik:1984vvp,Sklyanin:1988yz,Fring:1993mp,Kulish:1992qb}, and the algebra reduces to the pure reflection case
\begin{equation}
A_i(\theta) B_\alpha = K_{i\alpha}^{j\beta}(\theta) A_j(-\theta) B_\beta,
\end{equation}
where $K$ is the integrable boundary reflection matrix. The constraints derived for defects naturally reduce to the corresponding boundary conditions in this limit.

\subsection{Asymptotic Bethe ansatz equations} 
By solving the defect Yang-Baxter equations together with the bulk $S$-matrix and imposing suitable analyticity conditions, the reflection and transmission matrices can be determined. These quantities serve as essential inputs for evaluating physical observables associated with the defect. In particular, we employ the thermodynamic Bethe ansatz (TBA) \cite{PhysRev.147.303,Yang:1968rm,Zamolodchikov:1989cf} to compute the defect $g$-function. Within this framework, a multi-particle system in a large but finite volume is considered. In such a setting, the rapidities of particles are not arbitrary; they must satisfy the asymptotic Bethe ansatz equations (BAEs), which act as quantization conditions. In what follows, we present the derivation of the asymptotic BAEs in the presence of a defect.

Unless otherwise stated, the Bethe ansatz derivations below are restricted to scalar diagonal bulk scattering with a single particle species and no defect internal degrees of freedom. The formulas involving simultaneous reflection and transmission assume scalar parity-invariant non-topological defects in generalized free theories. More general non-diagonal defects require a separate diagonalization problem.

To illustrate the method, we first re-derive the simplest case: a periodic system with diagonal, pure bulk scattering. We introduce a test particle $A_{i}(\theta_i)$ into a system of length $L$ from the left, with the initial state given by
\begin{equation}
\mathbb{A}_{\{\mu\}}:=A_{\mu_{1}}(\theta_{\mu_{1}})\ldots A_{\mu_{N}}(\theta_{\mu_{N}})\ .
\end{equation}
Commuting the test particle through the state yields
\begin{equation}
A_{i}(\theta_i) \mathbb{A}_{\{\mu\}} = \prod_{j=1}^{N} S_{i \mu_{j}} (\theta_{i\mu_{j}}) \mathbb{A}_{\{\mu\}} A_{i}(\theta_i).
\end{equation}
Note that the rapidity representation does not explicitly encode any position space information. Since the vacuum state $|0\rangle$ is translationally invariant, the periodic boundary condition can be imposed through $\mathbb{A}_{\{\mu\}} A_{i}(\theta_i) $ as follows
\begin{equation}
\mathbb{A}_{\{\mu\}} A_{i}(\theta_i) = A_{i}(\theta_i) \mathbb{A}_{\{\mu\}} e^{ i L m_i \sinh \theta_i }.
\end{equation}
Requiring this relation to hold for all $A_{i}(\theta_i) \mathbb{A}_{\{\mu\}}$ leads to the BAE for periodic, diagonal IQFTs
\begin{equation}
e^{ i L m_i \sinh \theta_i } \prod_{j=1}^{N} S_{i \mu_{j}}(\theta_{i\mu_{j}}) =1.
\end{equation}

\paragraph{Topological defect.} The inclusion of a topological defect $D_{\alpha}$ (purely transmissive) is straightforward. The BAE becomes
\begin{equation}\label{eq:BAE_topo}
e^{ i L m_i \sinh \theta_i } T(\theta_{i}, \gamma) \prod_{j=1}^{N} S_{i \mu_{j}}(\theta_{i\mu_{j}}) =1,
\end{equation}
where $T(\theta_{i}, \gamma)$ is the transmission amplitude and $\gamma$ represents the defect parameter. For $N_{d}$ topological defects, this generalizes to
\begin{equation}
e^{ i L m_i \sinh \theta_i } \prod_{k=1}^{N_{d}}T(\theta_{i}, \gamma_{k}) \prod_{j=1}^{N} S_{i \mu_{j}}(\theta_{i\mu_{j}}) =1,
\end{equation}
where $\gamma_{k}$ is the corresponding defect parameter.

\paragraph{{Non-topological defect}.}
The derivation of the BAE for systems involving non-topological defects is more nontrivial. Previous approaches relied on a direct diagonalization of the ZF algebra \cite{Castro-Alvaredo:2002ulw}. Here, we give an alternative derivation that offers a clearer physical picture.

We focus on the parity-invariant non-topological defects in generalized free theories. The ZF algebra reads
\begin{align}
A_i(\theta) D_\alpha &= R_{i\alpha}(\theta) A_i(-\theta) D_\alpha + T_{i\alpha}(\theta) D_\alpha A_i(\theta), \label{eq:diagonal_defect_left}\\
D_\alpha A_i(-\theta)  &= R_{i\alpha}(\theta) D_\alpha A_i(\theta)  + T_{i\alpha}(\theta) A_i(-\theta) D_\alpha. \label{eq:diagonal_defect_right}
\end{align}
Consider an initial state created by the operator
\begin{equation}
\hat{\mathbb{A}}_{k,\alpha}^{\{\mu\}}:= A_{\mu_{1}}(\theta_{\mu_{1}})\ldots A_{\mu_{k}}(\theta_{\mu_{k}})D_{\alpha}\ldots A_{\mu_{N}}(\theta_{\mu_{N}}).
\end{equation}
Standard approaches begin with $A_i(\theta) \hat{\mathbb{A}}_{k,\alpha}^{\{\mu\}}$ and $\hat{\mathbb{A}}_{k,\alpha}^{\{\mu\}} A_i(\theta)$, eliminating the term containing $A_i(-\theta) \hat{\mathbb{A}}_{k,\alpha}^{\{\mu\}}$ to derive the BAE.

\begin{figure}[h!]
\centering
\includegraphics[scale=0.5]{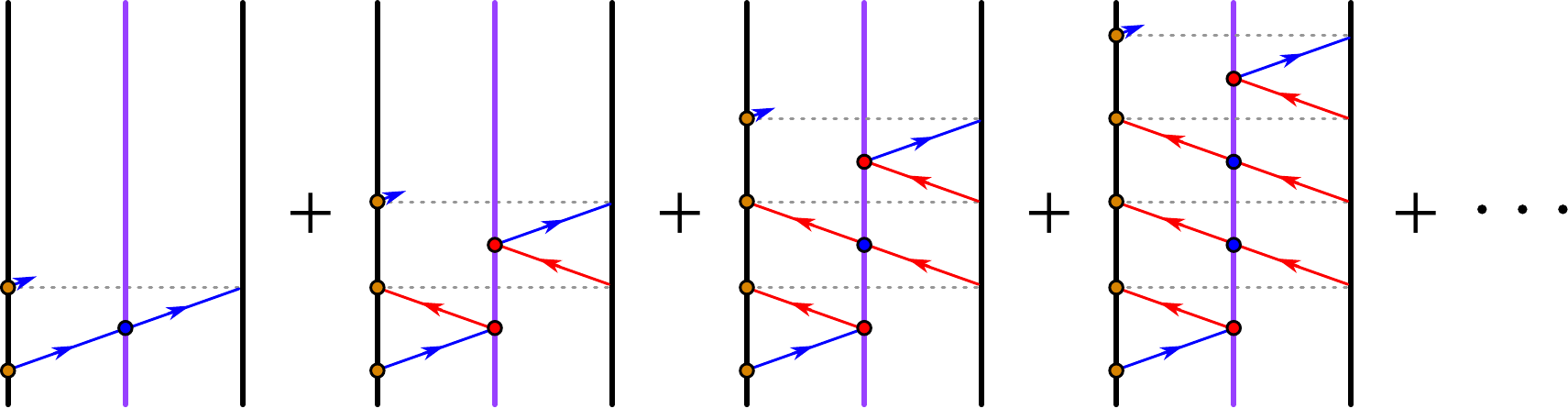}
\caption{Schematic illustrations of the terms in Bethe ansatz equations. The two black lines are identified and the magenta line indicates the defect line located at $y=0$. Blue and red lines represent right-moving and left-moving trajectories, respectively. Orange dots mark the initial and final position. Red dots indicate reflections off the defect, while blue dots indicate transmissions through the defect.}
\label{fig:BAE_non_topo}
\end{figure}
We provide a scattering interpretation of the ZF algebra. As illustrated in Fig.~\ref{fig:BAE_non_topo}, upon first encountering the defect line, a right-moving particle either transmits or reflects. The transmitted path closes the trajectory and contributes the usual phase $e^{i L m_i\sinh\theta_i}$. The reflected path winds around the circle and re-encounters the defect, generating an infinite multiple-scattering expansion.

The resulting contributions naturally group into the three classes depicted in Fig.~\ref{fig:BAE_non_topo}: (i) pure transmission; (ii) boundary-like pure reflection; and (iii) mixed sequences with reflections only at the first and last defect encounters and an arbitrary number of intermediate transmissions, which resum to a geometric series as $n\to\infty$.

Performing this calculation, the non-diagonal part vanishes, while the summed infinite series yields a finite contribution. This completes the diagonalization procedure, resulting in the BAE
\begin{equation}
e^{i L m_{i} \sinh \theta_i} \left( T_{i\alpha}(\theta_i,\gamma) \pm R_{i\alpha}(\theta_i,\gamma) \right) \prod_{l=1}^{N} S_{i\mu_l} = 1.
\end{equation}
Detailed derivations are provided in Appendix \ref{app:reduce}. This result agrees with the existing literature \cite{Castro-Alvaredo:2002ulw}. From the derivation, we see that the process involves a coherent superposition of left-moving and right-moving scattering modes. Therefore, the two signs do not correspond to purely left-moving or purely right-moving cases. In the limit $R_{i \alpha} \to 0$, we recover the purely transmissive case given in Eq.~\eqref{eq:BAE_topo}. For a purely transmissive defect that is not parity invariant, the right-moving mode instead obeys
\begin{equation}
e^{i L m_{i} \sinh \theta_i}  \tilde{T}_{i\alpha}(\theta_i,\gamma)  \prod_{l=1}^{N} S_{\mu_l i} = 1,
\end{equation}
which corresponds precisely to the pure right-moving mode for topological defects. A generic non-parity-invariant defect with both reflection and transmission would require diagonalizing the full matrix involving $T,\tilde T,R$ and $\tilde R$.

It should be emphasized that, to reduce this equation to the boundary case, one cannot simply take the limit of vanishing transmission amplitude. Instead, one should start from the original quadratic BAE and let $T_i^\alpha(\theta)\to 0$. In the parity-invariant scalar case this yields
\begin{equation}
e^{2 i L m_{i} \sinh \theta_i} R_{i\alpha}(\theta_i) R_{i\alpha}(\theta_i)\prod_{l=1}^{N} S_{i\mu_l} S_{\mu_{l}i} =1\,.
\end{equation}
This is precisely the boundary Bethe ansatz equation. 

\paragraph{Direct diagonalization}
{Building on the result above, we now present an alternative derivation based on a modified state vector in the ZF-algebra basis} 
\begin{equation}
\begin{split}
\begin{pmatrix}
A_{i}(\theta_i) \hat{\mathbb{A}}_{k,\alpha}^{\{\mu\}} \\
\hat{\mathbb{A}}_{k,\alpha}^{\{\mu\}} A_{i}(-\theta_i)
\end{pmatrix}
=&\begin{pmatrix}
T_{i \alpha}(\theta_i) \prod_{l=1}^{N} S_{i\mu_l}  & R_{i \alpha}(\theta_i)  \prod_{l=1}^{k} S_{i\mu_l} S_{\mu_{l}i} \\
R_{i \alpha}(\theta_i) \prod_{l=k+1}^N S_{\mu_{l}i} S_{i\mu_l}  & T_{i \alpha}(\theta_i) \prod_{l=1}^N S_{\mu_{l}i} 
\end{pmatrix}\\
&\times \begin{pmatrix}
\hat{\mathbb{A}}_{k,\alpha}^{\{\mu\}} A_{i}(\theta_i) \\
A_{i}(-\theta_i) \hat{\mathbb{A}}_{k,\alpha}^{\{\mu\}}
\end{pmatrix},
\end{split}
\end{equation}
Although this construction differs from that of \cite{Castro-Alvaredo:2002ulw}, it leads to the same quantization condition. Imposing the periodic boundary condition
\begin{equation}
\begin{split}
\begin{pmatrix}
A_{i}(\theta_i) \hat{\mathbb{A}}_{k,\alpha}^{\{\mu\}} \\
\hat{\mathbb{A}}_{k,\alpha}^{\{\mu\}} A_{i}(-\theta_i)
\end{pmatrix}
=&\begin{pmatrix}
T_{i \alpha}(\theta_i) \prod_{l=1}^{N} S_{i\mu_l}  & R_{i \alpha}(\theta_i)  \prod_{l=1}^{k} S_{i\mu_l} S_{\mu_{l}i} \\
R_{i \alpha}(\theta_i) \prod_{l=k+1}^N S_{\mu_{l}i} S_{i\mu_l}  & T_{i \alpha}(\theta_i) \prod_{l=1}^N S_{\mu_{l}i} 
\end{pmatrix}\\
&\times e^{i L m_i \sinh \theta_i} \begin{pmatrix}
A_{i}(\theta_i) \hat{\mathbb{A}}_{k,\alpha}^{\{\mu\}} \\
\hat{\mathbb{A}}_{k,\alpha}^{\{\mu\}} A_{i}(-\theta_i)
\end{pmatrix},
\end{split}
\end{equation}
we can directly diagonalize the matrix on the right-hand side to obtain the eigenvalues
\begin{equation}
\begin{split}
e^{i L m_i \sinh \theta_i}
\begin{pmatrix}
\left(T_{i \alpha}(\theta_i) - R_{i \alpha}(\theta_i)\right) \prod_{l=1}^{N} S_{i\mu_l}  & 0 \\
0  & \left(T_{i \alpha}(\theta_i) + R_{i \alpha}(\theta_i)\right) \prod_{l=1}^N S_{\mu_{l}i} 
\end{pmatrix}\ ,
\end{split}
\end{equation}
which reproduces the result derived previously.

\paragraph{{Multiple defects.}}
To study fusion, we need to consider the asymptotic BAE with multiple defect insertions. More specifically, we consider two defect insertions, with one defect located at the origin and the other at $y=a$. In the BAE, the defect away from the origin picks up an extra phase \cite{Castro-Alvaredo:2002ulw}, leading to
\begin{equation}
R_{i\alpha}(\theta,a) = R_{i\alpha}(\theta) e^{-2i ma \sinh \theta}, \quad \tilde{R}_{i\alpha}(\theta,a) = \tilde{R}_{i\alpha}(\theta) e^{2i ma \sinh \theta}
\end{equation}
The corresponding initial state is defined by
\begin{equation}\label{eq:ZF_unfuse_state}
\hat{\mathbb{A}}_{k_1, \alpha, k_2,\beta }^{\{\mu\}} := A_{\mu_{1}}\left(\theta_{\mu_{1}}\right) \ldots A_{\mu_{k_1}}(\theta_{\mu_{k_1}}) D_{\alpha} \ldots A_{\mu_{k_2}}(\theta_{\mu_{k_2}})  D_{\beta} \ldots A_{\mu_{N}}\left(\theta_{\mu_{N}}\right).
\end{equation}
Using the ZF algebra in the scalar parity-invariant case, we obtain
\begin{equation}
\begin{split}
A_{i}(\theta_i) \hat{\mathbb{A}}_{k_1, \alpha, k_2,\beta }^{\{\mu\}} &= \prod_{k=1}^{k_1} S_{i\mu_{k}} S_{\mu_{k}i} \mathcal{R}_{\alpha\beta}(\theta_i,a) A_{i}(-\theta_i) \hat{\mathbb{A}}_{k_1, \alpha, k_2,\beta }^{\{\mu\}} \\
&+ \prod_{k=1}^{N} S_{i\mu_{k}}   \mathcal{T}_{\alpha\beta}(\theta_i,a)  \hat{\mathbb{A}}_{k_1, \alpha, k_2,\beta }^{\{\mu\}} A_{i}(\theta_i),
\end{split}
\end{equation}
where we have defined
\begin{align}
\mathcal{R}_{\alpha\beta}(\theta_i,a) &= R_{i\alpha}(\theta_i) +\frac{T_{i\alpha}(\theta_i)^2 R_{i\beta}(\theta_i ,a) }{ 1-R_{i\alpha}(\theta_i) R_{i\beta}(\theta_i,a)} ,\label{eq:ising_eff_R} \\
\mathcal{T}_{\alpha\beta}(\theta_i ,a) &=   \frac{ T_{i\alpha}(\theta_i) T_{i\beta}(\theta_i)  }{1-R_{i\alpha}(\theta_i) R_{i\beta}(\theta_i ,a) } \label{eq:ising_eff_T}\,.
\end{align}
Here the parameter $a$ in $R_{i\beta}(\theta_i ,a)$ denotes the distance between the two defects. These expressions agree with the joint condition for two scalar parity-invariant defects \cite{Delfino:1994nr,Konik:1996rc,Castro-Alvaredo:2002ulw}. For particles incident from the left, the corresponding phase shifts are encoded in $\mathcal{R}_{\alpha\beta}(\theta_i,a)$. The denominator in \eqref{eq:ising_eff_R} and \eqref{eq:ising_eff_T} results from summing over multiple scattering processes in which the particle repeatedly bounces between the two defects. For non-parity-invariant scalar defects the same construction goes through with the appropriate left/right transmission factors, such as $T,\tilde T$, and left/right reflection amplitudes. For particles incident from the right, we have
\begin{equation}
\begin{split}
\hat{\mathbb{A}}_{k_1, \alpha, k_2,\beta }^{\{\mu\}}  A_{i}(-\theta_i) &= \prod_{k=k_2+1}^{N} S_{\mu_k i}  S_{i \mu_k} \tilde{\mathcal{R}}_{\alpha\beta}(\theta_i ,a) \hat{\mathbb{A}}_{k_1, \alpha, k_2,\beta }^{\{\mu\}} A_{i}(\theta_i)\\
& + \prod_{k=1}^{N} S_{\mu_k i} \mathcal{T}_{\alpha\beta} (\theta_i, a) A_{i}(-\theta_i) \hat{\mathbb{A}}_{k_1, \alpha, k_2,\beta }^{\{\mu\}} ,
\end{split}
\end{equation}
where
\begin{equation}
\tilde{\mathcal{R}}_{\alpha\beta}(\theta_i, a) = \tilde{R}_{i\beta}(\theta_i ,a) +\frac{T_{i\beta}(\theta_i)^2 R_{i\alpha}(\theta_i)  }{ 1-R_{i\alpha}(\theta_i) R_{i\beta}(\theta_i, a) } \label{eq:ising_eff_Rt}.
\end{equation}
Following the diagonalization method outlined above, we obtain the BAE
\begin{equation}\label{eq:BAE_two_nontopo}
e^{i L m_i \sinh \theta_i} \left( \mathcal{T}_{\alpha\beta}(\theta_i, a) \pm \sqrt{\mathcal{R}_{\alpha\beta}(\theta_i, a) \tilde{\mathcal{R}}_{\alpha\beta} (\theta_i, a) } \right) \prod_{k=1}^{N} S_{i\mu_{k}} =1.
\end{equation}
In Section \ref{sec:one}, we will discuss the fusion of non-topological defects using this BAE. The square root is taken on the branch continuously connected to the single-defect parity eigenvalues in the limit where the second defect is removed. In the special case of two identical defects, the BAE reduces to
\begin{equation}\label{eq:BAE_same_nontopo}
e^{i L m_i \sinh \theta_i} \left( \mathcal{T}_{\alpha\alpha}(\theta_i, a) \pm e^{i  a m_i \sinh \theta_i} \mathcal{R}_{\alpha\alpha}(\theta_i, a)  \right) \prod_{k=1}^{N} S_{i\mu_{k}} =1.
\end{equation}
As we will see, the additional phase factor $e^{i a m_i \sinh \theta_i}$ modifies the properties of the system.

\paragraph{{Defect between boundaries.}}
Finally, we consider the configuration with a defect and two integrable boundaries. The integrable boundaries $B_a$ and $B_b$ are placed at $y=-L$ and $y=0$ respectively, while the defect is located at $y=-a$. The initial state then takes the form
\begin{equation*}
B_{a} A_{i}(\theta_i) \hat{\mathbb{A}}_{k,\alpha}^{\{\mu\}} B_{b}
\end{equation*}
Using a similar method, we obtain the following boundary-like BAE
\begin{equation}
e^{2 i L m_i \sinh \theta_i} K_{a}(\theta_i) \mathcal{R}_{\alpha, b}(\theta_i, a) \prod_{l=1}^{N} S_{i \mu_l} S_{\mu_l i} = 1,
\end{equation}
where
\begin{equation}
\begin{split}
\mathcal{R}_{\alpha,b}(\theta_i, a) &= e^{-2 i m_i a \sinh \theta_i} R_{i\alpha}(\theta_i, -a)  +  \frac{ T_{i\alpha}(\theta_i)^2 K_b(\theta_i)}{ 1- \tilde{R}_{i\alpha}(\theta_i, -a) K_{b}(\theta_i)} \\
& = R_{i\alpha}(\theta_i)  +  \frac{ T_{i\alpha}(\theta_i)^2 K_b(\theta_i)}{ 1- \tilde{R}_{i\alpha}(\theta_i, -a) K_{b}(\theta_i)}.
\end{split}
\end{equation}
Here we consider a particle scattering from left to right. The defect and the boundary $B_b$ can be treated as a composite effective boundary. Multiple reflections between the defect and the boundary are not negligible and they can significantly modify the reflection amplitude.

When the reflection coefficient of the defect vanishes, the defect becomes topological, yielding the BAE
\begin{equation}
e^{2 i m_i L \sinh \theta_i} K_{a}(\theta_i) T_{-,i \alpha}(\theta_{i}) K_{b}(\theta_i) T_{+,i \alpha}(\theta_{i}) \prod_{l=1}^{N} S_{i \mu_l} S_{\mu_l i} = 1.
\end{equation}
Conversely, when the transmission of the defect vanishes, it effectively becomes a boundary, leading to the BAE
\begin{equation}
e^{2 i (L-a) m_i  \sinh \theta_i} K_{a}(\theta_i) R_{i \alpha}(\theta_{i}, -a)  \prod_{l=1}^{k} S_{i \mu_l} S_{\mu_l i} = 1.
\end{equation}
This reflects the decoupling of the two systems: particles cannot penetrate the region of length $a$ on the opposite side. Similarly, particles scattering from the right of the defect lead to another decoupled system of length $a$.

\subsection{Defect $g$-function}

The scattering picture discussed above applies in the infinite volume. In the finite volume, an important characteristic of the defect is the defect $g$-function, a natural defect analogue of the Affleck-Ludwig $g$-function \cite{Affleck:1991tk} for the boundary case. More precisely, it gives the non-extensive, $O(1)$ contribution of the defect to the finite-size free energy, or equivalently the defect entropy, and hence measures the effective degrees of freedom localized on the defect. The defect $g$-function therefore provides a finite volume probe complementary to the scattering data: it is sensitive to finite-size effects and allows one to track how the localized defect degrees of freedom evolve under changes of scale and under fusion.

\subsubsection{Definition}
\begin{figure}[tbp]
    \centering
    \begin{subfigure}[b]{0.49\linewidth}
        \centering
        \includegraphics[width=\textwidth]{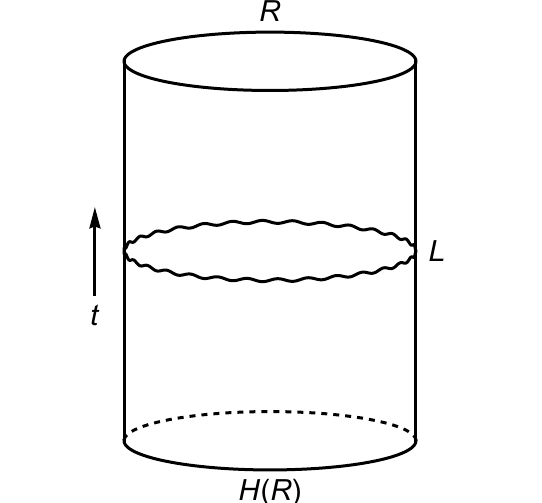}
        \caption{Closed channel} 
        \label{fig:closed}
    \end{subfigure}
    \hfill 
    \begin{subfigure}[b]{0.49\linewidth}
        \centering
        \includegraphics[width=\textwidth]{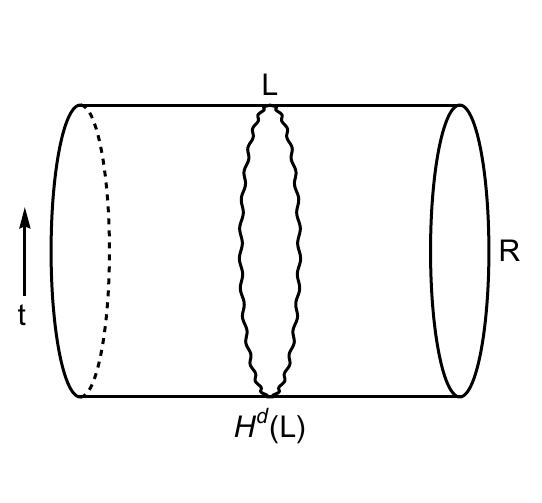}
        \caption{Open channel}
        \label{fig:open}
    \end{subfigure}    
    \caption{Schematic representation of the two channels for the {cylinder partition function}. The dashed line represents the defect line parallel to the $L$ direction, and the arrows indicate the direction of Euclidean time evolution. $R$ denotes the periodic circumference.}
\end{figure}

We first define the defect $g$-function. Consider an IQFT on a Euclidean cylinder with a compact spatial direction with circumference $R$, as shown in Fig.~\ref{fig:closed}.

\paragraph{Periodic case}
We first consider a single integrable defect and impose periodic boundary conditions along the temporal direction, with circumference $L$. In this special case, the cylinder becomes a torus. The partition function can then be evaluated in two equivalent quantization channels, as shown in Fig. \ref{fig:closed} and Fig. \ref{fig:open} respectively. We shall refer to these two channels as `closed' and `open' channels in what follows.

In the closed channel, the defect corresponds to the insertion of a defect operator $\mathcal{D}$ acting on the Hilbert space. The partition function is given by \cite{Bajnok:2004jd,Bajnok:2007jg}
\begin{equation}\label{eq:pfclosed}
Z(R,L) = \Tr \left( \mathcal{D} e^{-H(R)L} \right)
= \sum_{i} \bra{i}\mathcal{D}\ket{i} e^{-E_i(R) L},
\end{equation}
where $H(R)$ denotes the Hamiltonian on the circle $S^1$ of length $R$, and $E_i(R)$ are the corresponding eigenvalues. The matrix elements of the defect operator depend on the system size $R$. In the limit $mL \gg 1$, the partition function is dominated by the ground-state contribution
\begin{equation}
Z(R,L) \approx \bra{0}\mathcal{D}\ket{0} e^{-E_0(R) L}.
\end{equation}
The ground-state energy $E_0(R)$ contains only the bulk contribution up to exponential corrections
\begin{equation}
E_0(R) = \epsilon R + \mathcal{O}(e^{-mR}) ,
\end{equation}
where $\epsilon$ is the bulk energy density.

In the open channel, by contrast, $L$ is regarded as the volume of the spatial direction and $R$ as the Euclidean time. The system is then quantized on a circle of length $L$, with the defect encoded in a modified Hamiltonian. The partition function reads
\begin{equation}\label{eq:pfopen}
Z(R,L) = \Tr e^{-H^d(L)R} = \sum_{i} e^{-E^d_i(L) R} ,
\end{equation}
Here, $H^d(L)$ is the Hamiltonian of the system of length $L$ in the presence of the defect, and $E^d_i(L)$ are the corresponding eigenvalues. In the limit $mR \gg 1$, the partition function is dominated by the ground state of the defect Hamiltonian
\begin{equation}
Z(R,L) \approx e^{-E^d_0(L) R} .
\end{equation}
The ground-state energy $E^d_0(L)$ contains both the bulk contribution and the defect contribution
\begin{equation}
E^d_0(L) = \epsilon L + f_d + \mathcal{O}(e^{-m L}),
\end{equation}
where $f_d$ is the non-extensive defect energy, which can also be interpreted as the defect tension.

Since the closed and open channels describe the same physical system, the partition functions in Eq.~\eqref{eq:pfclosed} and Eq.~\eqref{eq:pfopen} must agree. In the regime $mR \gg 1$ and $mL \gg 1$, matching the leading terms relates the vacuum expectation value (VEV) of the defect operator to the defect $g$-function
\begin{equation}
\expval{\mathcal{D}} = e^{-f_d R} \left[1+ \mathcal{O}(e^{-mR})\right],
\end{equation}
where $\expval{\mathcal{D}} \equiv \bra{0}\mathcal{D}\ket{0}$. We define the defect $g$-function, denoted by $g_d(mR)$, through
\begin{equation}
\expval{\mathcal{D}} = e^{-f_d R} g_d(mR).
\end{equation}

For later convenience, it is useful to introduce the subtracted partition functions \cite{Pozsgay:2010tv} for the bulk system and the defect system,
\begin{align}
Z_0(R,L) &= \sum_{i} e^{-( E_i(L) - \epsilon L)R},\\
Z_1(R,L) &= \sum_{i} e^{-( E_i^{d}(L) -f_d - \epsilon L)R}.
\end{align}
The defect $g$-function can then be written as
\begin{equation}\label{eq:cal-gfunc}
\log g_d(mR) = \lim_{L \to \infty} \left[\log Z_1(R,L)- \log Z_0(R,L) \right].
\end{equation}
The partition functions $Z_1$ and $Z_0$ can be evaluated by TBA.

\paragraph{Multiple defects.} We now extend the above definition to the case of multiple defects, which will be needed for fusion. For a system with two defects, labeled $a$ and $b$, we place defect $a$ at origin and defect $b$ at $y$, with $0<y<L$. By translation invariance only the separation $y$ is relevant. In the open channel, the partition function becomes
\begin{equation}
Z(R,L;y) = \Tr e^{-H^{d_a,d_b}(L;y)R} \approx e^{-E_0^{d_a,d_b}(L;y) R},
\end{equation}
where $H^{d_a,d_b}(L;y)$ denotes the Hamiltonian on the circle of length $L$ in the presence of the two defects. If the defects are well separated, namely $my \gg 1$ and $m(L-y)\gg 1$, then
\begin{equation}
E_0^{d_a,d_b}(L;y) = \epsilon L + f_{d,a} + f_{d,b} + \mathcal{O}(e^{-m y})+\mathcal{O}(e^{-m(L-y)}).
\end{equation}
In the closed channel, this corresponds to inserting defect operators at Euclidean times $0$ and $y$, yielding
\begin{equation}
\begin{split}
Z(R,L;y) &= \Tr \left( e^{-H(R)(L-y)} \mathcal{D}_b e^{-H(R)y} \mathcal{D}_a \right) \\
&= \sum_{i} e^{-E_i(R)(L-y)} \sum_{j} \bra{i}\mathcal{D}_b\ket{j} \bra{j}\mathcal{D}_a\ket{i} e^{-E_j(R)y}\\
&\approx \expval{\mathcal{D}_b(y)\mathcal{D}_a(0)} e^{-E_0(R) L}.
\end{split}
\end{equation}
Accordingly, the $g$-function for the two-defect system can be defined via
\begin{equation}
\expval{\mathcal{D}_b(y)\mathcal{D}_a(0)} = e^{-(f_{d,a} + f_{d,b})R} g_{d,ab}(mR,my).
\end{equation}
This quantity can again be computed from Eq.~\eqref{eq:cal-gfunc}, provided that $Z_1(R,L)$ is replaced by the partition function associated with the Hamiltonian in the presence of two defects,
\begin{align}
Z_1(R,L;y) &= \sum_{i} e^{-( E_i^{d_a,d_b}(L;y) - f_{d,a} - f_{d,b} - \epsilon L)R}.
\end{align}

At this stage it is important to distinguish between topological and non-topological defects in IQFTs. This distinction is most transparent in the closed channel, where the VEV of the product of defect operators satisfies the decomposition property
\begin{equation}
\expval{\mathcal{D}_b(y)\mathcal{D}_a(0)} = 
\begin{cases}
\bra{0}\mathcal{D}_b\ket{0} \bra{0}\mathcal{D}_a \ket{0}, & \text{for topological defects}, \\
\sum_{j} e^{-(E_j(R)-E_0(R))y}\bra{0}\mathcal{D}_b\ket{j} \bra{j}\mathcal{D}_a \ket{0}, & \text{for non-topological defects}.
\end{cases}
\end{equation}
For topological defects, the factorization arises because the defect operators commute with the translation operator and effectively have diagonal matrix elements in the rapidity basis \cite{Jiang:2017qhn}. Consequently, the $g$-function for two unfused topological defects factorizes into the product of the individual contributions
\begin{align}
\expval{\mathcal{D}_b} \expval{\mathcal{D}_a} &= e^{-f_{d,a}R } g_{d,a}(mR) \cdot e^{- f_{d,b}R} g_{d,b}(mR),\\
\log g_{d,ab}(mR) &= \log g_{d,a}(mR) + \log g_{d,b}(mR).
\end{align}
This implies that for topological defects, the total contribution to the $g$-function is obtained simply by multiplying $g$-functions of the individual defects, and the dependence on $y$ drops out. Non-topological defects do not enjoy this factorization property because of the non-trivial sum over intermediate states $\ket{j}$ \cite{Jiang:2017tyi}. 

\paragraph{Defect between boundaries.} We now extend our analysis to the case where a defect is placed between two integrable boundaries. This setup is necessary for defining the fusion of defects with boundaries.

In the closed channel, the system evolves between two boundary states \cite{Ghoshal:1993tm}, $\ket{B_a}$ and $\ket{B_b}$, rather than under periodic boundary conditions. The defect operator $\mathcal{D}$ is inserted between these boundary states, and the partition function is expressed as the matrix element \cite{LeClair:1995uf,Dorey:2004xk,Pozsgay:2010tv}
\begin{equation}
Z(R,L) = \bra{B_a} \mathcal{D} e^{-H(R) L} \ket{B_b}.
\end{equation}
After inserting complete sets of bulk eigenstates $\ket{i}$ and $\ket{j}$, and assuming the defect is topological, in the $L\gg1$ limit we obtain
\begin{equation}
Z(R,L) 
\approx (G_a^{0}(mR))^{\ast} \expval{\cal D} G_b^{0}(mR) e^{-E_0(R)L}\,,
\end{equation}
where $G_j^0 = \braket{0}{B_j}$ $(j=a,b)$ and $\expval{\cal D} = \bra{0}\mathcal{D}\ket{0}$.

In the open channel, the system is defined on an interval of length $L$ with integrable boundary conditions labeled by $a$ and $b$ at the two ends, together with a defect $\cal D$ inserted at an intermediate position. The ground-state energy $E_{d}^0(L)$ now contains non-extensive contributions from both the boundaries and the defect. In the limit $mR \gg 1$, the partition function defines a mixed defect-boundary $g$-function, denoted by $g_{d,b}(mR)$, through
\begin{equation}
(G_{a}^{0})^{*} \expval{\cal D} G_{b}^{0} = e^{-R (f_d + f_{b,a} + f_{b,b})} g_{d,b}(mR).
\end{equation}
Here, $f_{b,a}$ and $f_{b,b}$ are the boundary free energies. For non-topological defects, excited states $\ket{i}$ also contribute and the left-hand side would involve a sum of the form $(G_{a}^{i})^{*} D_{i,0} G_{b}^{0}$, where $D_{i,0} \equiv \bra{i}\mathcal{D}\ket{0}$. 

Recalling that the boundary $g$-functions are defined by \cite{LeClair:1995uf,Dorey:2004xk,Pozsgay:2010tv}
\begin{equation}
G_a^0 = e^{-R f_{b,a}} g_{b,a}(mR),
\end{equation}
we find that the total $g$-function decomposes into three independent pieces
\begin{equation}
\log g_{d,b}(mR) = \log g_{d}(mR) + \log g_{b,a}(mR) + \log g_{b,b}(mR).
\end{equation}
This additivity of the logarithmic $g$-function is a characteristic feature of the topological defects. In particular, when a topological defect is present together with boundaries but remains spatially separated from them, its contribution to the boundary entropy is additive. In the next subsections, we develop a general method for computing the defect $g$-function in different situations.

\subsubsection{TBA approach to defect $g$-function}

{In the closed channel, the defect operator is inserted in the partition function, a formulation that is particularly natural for topological defects \cite{Bajnok:2004jd,Bajnok:2007jg}. Alternatively, one may work in the open channel and extract the $g$-function, following \cite{Pozsgay:2010tv}. In what follows, we adopt the latter channel and generalize it to systems with defects: the partition function is considered at finite $R$, whereas the Bethe-Yang equations describe the large-volume quantization in the open channel.} 

Assuming diagonal bulk scattering with two-body scattering matrix $S(\theta)$, we now consider the {open-channel large-volume} quantization of $N$ particles in a system of spatial size $L$ in the presence of defects or boundaries. Generalizing the BAEs introduced above, the corresponding Bethe-Yang equation can be written as
\begin{equation}\label{eq:gen_BAE}
e^{i \nu L p(\theta_{j})} P(\theta_j) \prod_{k\ne j} \mathcal{S}(\theta_j,\theta_k)=1, \quad j=1,\dots,N,
\end{equation}
for a relativistic massive field theory with a single particle species and no internal degrees of freedom. Here, $p(\theta_{j}) = m \sinh \theta_j$ is the momentum of the particle, and $\nu$ is a geometric factor encoding the boundary conditions. The factor $P(\theta_j)$ collects the contributions of defects or boundaries, while $\mathcal{S}(\theta_j,\theta_k)$ denotes the relevant bulk scattering amplitude. Their explicit forms depend on the specific geometry and on the defect or boundary configuration.

Taking the logarithm of Eq.~\eqref{eq:gen_BAE}, we obtain the counting function in terms of the quantum numbers $I_j$
\begin{equation}
Q(\theta_j) = \nu L p(\theta_j) - i \log P(\theta_j) - i \sum_{k \neq j} \log \mathcal{S}(\theta_j, \theta_k) = 2\pi I_j, \quad I_j \in \mathbb{Z}.
\end{equation}
In the thermodynamic limit, defined by $L \to \infty$ and $N \to \infty$ with the ratio $N/L$ fixed, the Bethe roots become continuously distributed. We therefore introduce the particle density $\rho_{p}(\theta)$ and the hole density $\rho_h(\theta)$, so that the total density is
\begin{equation}
\frac{1}{L}\frac{\dd I_{j}}{\dd \theta}
=\frac{1}{2\pi L}\frac{\dd Q(\theta)}{\dd \theta}
= \rho_{p}(\theta) + \rho_h(\theta) = \rho(\theta).
\end{equation}
Differentiating the counting function with respect to $\theta$ yields the density constraint
\begin{equation}\label{eq:density_constraint}
\rho_{p}(\theta) + \rho_h(\theta) = \frac{1}{2\pi} \sigma(\theta) + \int \frac{\dd\theta'}{2\pi} \varphi(\theta, \theta') \rho_{p}(\theta'),
\end{equation}
where the kernel $\varphi$ and the source term $\sigma$ are defined by
\begin{equation}
\varphi(\theta, \theta') = -i \frac{\partial}{\partial \theta} \log \mathcal{S}(\theta, \theta'), \quad \sigma(\theta) = \nu \frac{\dd p(\theta)}{\dd\theta} + \Theta(\theta),
\end{equation}
with the defect or boundary contribution
\begin{equation}
\Theta(\theta) = -i \frac{1}{L} \frac{\dd}{\dd \theta}
\log \Big( P(\theta)\,\mathcal{S}_{\rm sub}(\theta) \Big).
\end{equation}
Here $\mathcal{S}_{\rm sub}(\theta)$ denotes the conventional self-interaction subtraction. The precise choice of this factor depends on the geometry and on the Bethe-Yang convention: it is trivial for purely transmitting periodic defects, while in boundary-type quantization it is usually related to the $S(2\theta)$ self-scattering subtraction.

The partition function is first defined for finite but large spatial size $L$, where the states are labeled by discrete Bethe quantum numbers. One then takes the thermodynamic limit, $N\to\infty, L\to\infty$ with $N/L$ fixed, and replaces the sum over quantum numbers by a functional integral over rapidity densities. In this limit, the leading contribution is determined by minimizing the free energy. However, passing from the quantum-number measure to the rapidity-density measure introduces a non-trivial Jacobian \cite{Pozsgay:2010tv}. 

We first summarize the saddle-point result \cite{Pozsgay:2010tv}. The free energy depends on both the energy and the entropy, which are naturally expressed in rapidity space. The number of micro-canonical configurations is
\begin{equation}
\Omega= \prod_\theta \omega(\rho_p(\theta)),
\end{equation}
where the combinatorial factor is
\begin{equation}
\omega=
\begin{pmatrix}
L\Delta \theta \rho(\theta)\\   L\Delta \theta \rho_p(\theta)
\end{pmatrix}
=
\frac{\big(L\Delta \theta \rho(\theta)\big)!}
{\big(L\Delta \theta \rho_p(\theta)\big)!\big(L\Delta \theta \rho_h(\theta)\big)!}.
\end{equation}
In the thermodynamic limit, this expression simplifies via Stirling's approximation
\begin{equation}
\log  \omega=L\Delta \theta
s(\rho(\theta))+\varsigma(\rho(\theta))+\dots\ ,
\end{equation}
where the entropy density is
\begin{equation}
s(\rho(\theta))=\rho(\theta)\log(\rho(\theta)) -\rho_p(\theta)\log(\rho_p(\theta))-\rho_h(\theta)\log(\rho_h(\theta))
\end{equation}
and the subleading correction relevant to the Gaussian fluctuations is
\begin{equation}
\varsigma(\rho(\theta))=-\frac{1}{2}\log(2\pi L\Delta\theta)+
\frac{1}{2}\log\frac{\rho(\theta)}{\rho_p(\theta)\rho_h(\theta)}  .
\end{equation}
The free energy functional is
\begin{equation}
\label{Ffunct}
F[\rho(\theta)]=L\sum_\theta \Big(e(\theta)\rho_p(\theta)-Ts(\rho(\theta))\Big)\Delta \theta.
\end{equation}
Minimizing this functional subject to the density constraint yields the thermodynamic equilibrium condition for the pseudo-energy $\varepsilon(\theta) \equiv \log (\rho_h(\theta)/\rho_{p}(\theta))$
\begin{equation}
\label{TBAg}
\beta e(\theta)=\varepsilon(\theta)+\int \frac{\dd\theta'}{2\pi}\  \varphi(\theta,\theta') \log\big(1+e^{-\varepsilon(\theta')}\big),
\end{equation}
where $e(\theta) = m \cosh \theta$ is the single-particle energy and $\beta = 1/T = R$ is the inverse temperature. This leads to the minimized free energy
\begin{equation}
F_{\rm min}=-LT \int \frac{\dd \theta}{2\pi}  \sigma(\theta)\log\big(1+e^{-\varepsilon(\theta)}\big).
\end{equation}

The partition function contains not only the exponential of the minimized free energy but also a prefactor arising from the density of states in configuration space:
\begin{equation}
Z(R,L) = \mathcal{N} \exp\left( - \beta F_{\rm min} \right).
\end{equation}
The normalization factor $\mathcal{N}$, which contributes at order $\mathcal{O}(1)$, is given by the Fredholm determinant \cite{Pozsgay:2010tv} associated with the integral operator derived from the density constraint Eq.~\eqref{eq:density_constraint}
\begin{equation}
\mathcal{N} = \det \left( 1 - \hat{G}^{\prime} \right),
\end{equation}
where the operator $\hat{G}^\prime$ acts on a test function $f(\theta)$ as
\begin{equation}
\hat{G}^{\prime} \circ f (\theta) = \int \frac{\dd\theta'}{2\pi} \varphi^{\prime}(\theta, \theta') \frac{1}{1+e^{\varepsilon(\theta')}} f(\theta')
\end{equation}
and the kernel is
\begin{equation}
\varphi^{\prime}(\theta, \theta') = i \frac{\partial}{\partial \theta^{\prime}} \log \mathcal{S}(\theta, \theta').
\end{equation}
Including Gaussian fluctuations around the saddle-point yields the ratio of Fredholm determinants \cite{Woynarovich:2004gc}
\begin{equation}
Z(R,L) = \exp\left( - \beta F_{\rm min} \right) \frac{\det \left( 1 - \hat{G}^{\prime} \right)}{ \det \left( 1 - \hat{G} \right) },
\end{equation}
where
\begin{equation}
\hat{G} \circ f (\theta) = \int \frac{\dd\theta'}{2\pi} \varphi(\theta, \theta') \frac{1}{1+e^{\varepsilon(\theta')}} f(\theta').
\end{equation}

\paragraph{General formula for the $g$-function.}
We can now apply this formalism to the $g$-function. In the present subsection we first write the formula for the total finite contribution extracted from the ratio of the partition function $Z_1$ with defects or boundaries to the reference bulk partition function $Z_0$. This total contribution may include defect terms, boundary terms and the Fredholm-determinant contribution, depending on the geometry. Substituting the explicit expressions, we obtain
\begin{equation}
\log g_{\rm tot}(mR) = \int \frac{\dd\theta}{2\pi} L \Theta(\theta) \log \left( 1 + e^{-\varepsilon(\theta)} \right)  + \log \frac{\det (1 - \hat{G}^{\prime})}{\det (1 - \hat{G})} .
\end{equation}
The determinant contribution appears with unit coefficient because $g_{\rm tot}$ is defined from the full partition-function ratio, before any further division by separately defined boundary $g$-functions. If one wants the contribution of a specific object, such as a pure defect contribution $g_d$ or a mixed defect-boundary contribution $g_{d,b}$, the appropriate boundary reference factors must be subtracted.

For a system with defects only, the scattering kernel $\mathcal{S}(\theta,\theta')$ typically coincides with the bulk one, $\mathcal{S}(\theta,\theta') = S(\theta-\theta')$. Hence $\hat{G}^{\prime}=\hat{G}$, and the ratio of Fredholm determinants is unity. In this case $g_{\rm tot}$ reduces to the defect contribution $g_d$. For boundary systems, by contrast, the scattering kernel is modified, so the determinant ratio becomes non-trivial and must be retained.

\subsubsection{Examples}

\paragraph{Sinh-Gordon theory}
As an illustrative example, we consider the topological defect in Sinh-Gordon theory, whose transmission amplitudes are \cite{Bajnok:2007jg}
\begin{equation}
T_{-}(\theta) = -i \frac{\sinh(\frac{\theta}{2} - \frac{i\pi}{4} + \frac{B\kappa}{2})}{\sinh(\frac{\theta}{2} + \frac{i\pi}{4} + \frac{B\kappa}{2})}, \quad 
T_{+}(\theta) = i \frac{\sinh(\frac{\theta}{2} - \frac{i\pi}{4} - \frac{B\kappa}{2})}{\sinh(\frac{\theta}{2} + \frac{i\pi}{4} - \frac{B\kappa}{2})},
\end{equation}
where $-$ and $+$ denote transmission through the topological defect from left to right and from right to left, respectively. The bulk $S$-matrix is \cite{Zamolodchikov:1995xk}
\begin{equation}
S(\theta) = \frac{\sinh \theta - i \sin B\pi}{\sinh \theta + i \sin B\pi},
\end{equation}
which has no poles in the physical strip. The parameter $B$ is related to the coupling constant $b$ by
\begin{equation}
B = \frac{b^2}{8\pi + b^2}.
\end{equation}
The corresponding BAE is
\begin{equation}
e^{ i L m \sinh \theta_i } T_-(\theta_{i}, \kappa) \prod_{j=1}^{N} S_{i \mu_{j}}(\theta_{i\mu_{j}}) =1, \quad e^{ i L m \sinh \theta_i } T_+(\theta_{i}, \kappa) \prod_{j=1}^{N} S_{\mu_{j} i}(\theta_{\mu_{j} i}) =1,
\end{equation}
and the defect $g$-function is
\begin{equation}
\log g_{d} = \int_{-\infty}^\infty \frac{\dd\theta}{2\pi} \left( -i \frac{\partial}{\partial \theta} \log T_{\pm}(\theta)\right) \log \left(1 + e^{- \varepsilon(\theta)}\right),
\end{equation}
where $\varepsilon(\theta)$ satisfy Eq.~\eqref{TBAg}. 

\begin{figure}[tbp]
    \centering
    \begin{subfigure}[b]{0.47\linewidth}
        \centering
        \includegraphics[width=\textwidth]{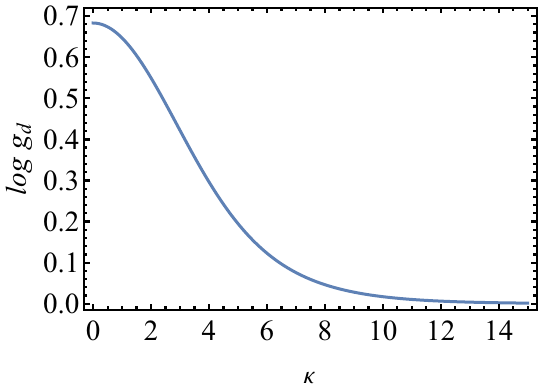}
        \caption{} 
        \label{fig:g_sinh-kappa}
    \end{subfigure}
    \hfill 
    \begin{subfigure}[b]{0.51\linewidth}
        \centering
        \includegraphics[width=\textwidth]{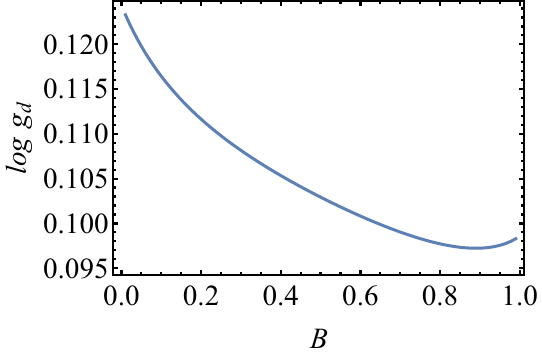}
        \caption{}
        \label{fig:g_sinh-b}
    \end{subfigure}    
    \caption{The function $\log g_{d}$ plotted against the defect parameter $\kappa$ in panel (a) and coupling constant $B$ in panel (b).}
    \label{fig:g_sinh}
\end{figure}

\paragraph{Numerical results}
We solve the TBA equation iteratively, and the two transmission directions give the same $g$-function. The results are shown in Fig.~\ref{fig:g_sinh}. In the IR limit, the pseudo-energy approaches $mR \cosh \theta$, so the logarithm of the $g$-function vanishes. 

The $g$-function is not monotonic when the bulk parameters are varied. This is allowed because the varying parameters are bulk couplings rather than defect parameters \cite{Green:2007wr,He:2024asy}. By contrast, the $g$-function decreases monotonically as the defect parameter $\kappa$ is increased in the range shown. This behavior is compatible with monotonicity expected along genuine defect RG flows \cite{Friedan:2003yc,Casini:2016fgb,Casini:2022bsu,Harper:2024aku}, although varying $\kappa$ should not automatically be identified with an RG trajectory without an explicit flow interpretation.

\paragraph{Ising field theory.}
Another example is the Ising field theory with a non-topological defect \cite{Delfino:1994nr,Delfino:1994nx}. The bulk $S$-matrix is trivial, $S(\theta)=-1$, while the transmission and reflection amplitudes of the defect are
\begin{equation}\label{eq:isingTR}
T(\theta, \chi) = \tilde{T}(\theta, \chi) =\frac{\cos \chi \sinh \theta}{\sinh \theta-i \sin \chi}, \quad 
R(\theta, \chi) = \tilde{R}(\theta, \chi) = i \frac{\sin \chi \cosh \theta}{\sinh \theta-i \sin \chi},
\end{equation}
where the parameter $\chi$ is related to the Ising defect coupling $g_{\rm I}$ by
\begin{equation}
\sin \chi = -\frac{4g_{\rm I}}{g_{\rm I}^2+4}.
\end{equation}
In the following we restrict to positive $g_{\rm I}$, since negative $g_{\rm I}$ leads to a bound state and lies outside the scope of the present discussion. When $g_{\rm I} = \pm 2$, the defect becomes purely reflective. For the composition law below we use the branch
\begin{equation}\label{eq:ising_chi_branch}
\chi=-2\arctan\dfrac{g_{\rm I}}{2}, \quad \chi \in [-\frac{\pi}{2},0],
\end{equation}
which resolves the two-to-one map in the sine parametrization. If the defect is located at a distance $a$ from the origin, the reflection amplitude becomes \cite{Delfino:1994nr,Castro-Alvaredo:2002ulw}
\begin{equation}\label{eq:reflectiona}
R(\theta, \chi , a) = \tilde{R}(\theta, \chi, -a) = i \frac{\sin \chi \cosh \theta}{\sinh \theta-i \sin \chi} e^{-2i a m \sinh\theta},
\end{equation}
while the transmission amplitudes remain unchanged. For simplicity, we place the defect at the origin.

{The two combinations $T(\theta)\pm R(\theta)$ have a direct interpretation in the diagonal basis of the defect scattering problem. Since the defect is parity invariant, scattering of particles incident from the two sides is described by the matrix}
\begin{equation*}
\begin{pmatrix}
T(\theta) & R(\theta)\\
R(\theta) & T(\theta)
\end{pmatrix},
\end{equation*}
{whose parity-even and parity-odd eigenvectors have eigenvalues $T(\theta)+R(\theta)$ and $T(\theta)-R(\theta)$, respectively. Thus the two signs label the two parity sectors of the diagonalized defect problem; they should not be identified with purely left-moving and right-moving sectors.}

For the Ising amplitudes, the assignment of the relevant parity sector is fixed by the strong--weak duality \cite{Delfino:1994nx}
\begin{equation}
T(\theta, 4/g_{\rm I}) = -T(\theta, g_{\rm I}), \quad R(\theta, 4/g_{\rm I}) = R(\theta, g_{\rm I}).
\end{equation}
It follows that
\begin{equation*}
T(\theta, 4/g_{\rm I})\pm R(\theta, 4/g_{\rm I})=-\left(T(\theta, g_{\rm I})\mp R(\theta, g_{\rm I})\right).
\end{equation*}
The duality therefore exchanges the two parity sectors, up to a rapidity-independent sign. Following a single duality orbit along the positive-coupling line then gives the $-$ sector for $0<g_{\rm I}<2$ and the $+$ sector for $g_{\rm I}>2$. The defect $g$-function is therefore
\begin{equation}
\log g_{d} =  
\begin{cases}
\mathcal{G}_{d}^{-} \quad \text{for}~ 0<g_{\rm I}<2\\
\mathcal{G}_{d}^{+} \quad \text{for}~ g_{\rm I}>2
\end{cases}
\end{equation}
with
\begin{equation}
\mathcal{G}_{d}^{\pm} = - i \int_{-\infty}^\infty \frac{\dd \theta}{2\pi} \left(  \frac{\partial}{\partial \theta} \log (T(\theta) \pm  R(\theta))\right) \log \left(1 + e^{-m R \cosh \theta}\right).
\end{equation}
Substituting the amplitudes from Eq.~\eqref{eq:isingTR}, we obtain
\begin{align}
\mathcal{G}_{\text{defect}}^{+} = -i \int_{-\infty}^\infty \frac{\dd \theta}{2\pi} \frac{\coth \left(\frac{1}{2} (\theta +i \chi )\right)\sin \chi}{\sin \chi +i \sinh \theta } \log \left(1 + e^{-m R \cosh \theta}\right) ,\\
\mathcal{G}_{\text{defect}}^{-} = -i \int_{-\infty}^\infty \frac{\dd \theta}{2\pi} \frac{\tanh \left(\frac{1}{2} (\theta - i \chi )\right)\sin \chi}{\sin \chi +i \sinh \theta } \log \left(1 + e^{-m R \cosh \theta}\right).
\end{align}

\begin{figure}[tbp]
    \centering
    \begin{subfigure}[b]{0.49\linewidth}
        \centering
        \includegraphics[width=\textwidth]{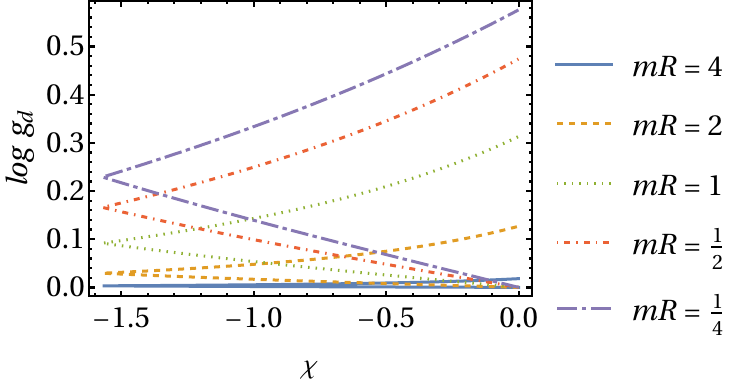}
        \caption{} 
        \label{fig:g_chi}
    \end{subfigure}
    \hfill 
    \begin{subfigure}[b]{0.49\linewidth}
        \centering
        \includegraphics[width=\textwidth]{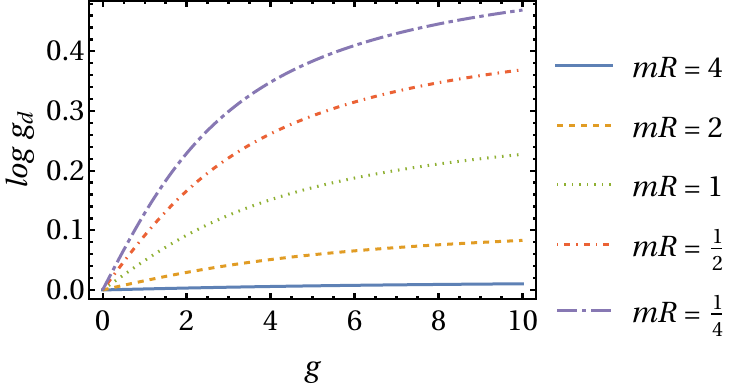}
        \caption{}
        \label{fig:g_g}
    \end{subfigure}    
    \caption{The function $\log g_{d}$ plotted against the defect parameter $\chi$ in panel (a) and the Ising defect coupling $g_{\rm I}$ in panel (b).}
\end{figure}

As shown in Fig.~\ref{fig:g_g}, this duality-compatible branch assignment gives a monotonically increasing $g$-function along the positive-coupling line. This is compatible with the expected direction of a defect RG flow, but the identification with an RG trajectory should be stated separately.

\paragraph{Comment on the dependence of $\chi$.}
A subtle issue arises in the limit $\chi \to 0$ of $\mathcal{G}_{d}^{+}$, since this limit does not commute with the integration. If the limit is taken before the integral is evaluated, the result vanishes, whereas the correct order of operations yields a non-trivial contribution. The reason is that, as $\chi \to 0$, a pole in the expression for $\mathcal{G}_{d}^{+}$ approaches the real axis. This produces an apparent discontinuity in $\mathcal{G}_{d}^{+}$ at $\chi = 0$, corresponding to the limit $g_{\rm I} \to \infty$ in $+$ branch.

The apparent non-smooth behavior in Fig.~\ref{fig:g_chi} is therefore entirely due to the choice of the folded $\chi$ variable. Although the algebraic expressions are simpler when written in terms of $\chi$, the sine relation between $g_{\rm I}$ and $\chi$ is not one-to-one. In contrast, the plot against $g_{\rm I}$ follows the physical positive-coupling line without this artificial folding.

The UV limit of the $g$-function is
\begin{equation}
\lim_{mR \to 0} \mathcal{G}_{d}^{\pm} =
\begin{cases}
-\frac{\log 2}{\pi} \chi  \quad \text{for } - \text{branch}\\
\frac{\log 2}{\pi} \chi +\log 2 \quad \text{for } + \text{branch}
\end{cases},
\end{equation}
whereas the IR limit $mR\to \infty$ is trivial, so that the logarithm of the $g$-function vanishes. It is worth noting that the parameter $\chi$ shifts the integration contour across a branch point, thereby moving the result onto the next Riemann sheet, from which the constant term $\log 2$ emerges.

\section{Fusion of defects}
\label{sec:one}

In this section, we study fusion of integrable defects, including defect-defect fusion and defect-boundary fusion. By fusion we mean the short-distance limit in which two separated objects are replaced by a single effective defect or boundary. Studying this limit is motivated by two related questions. First, it provides a concrete way to follow how the localized degrees of freedom carried by the defects combine, and whether the topological information encoded by the original configuration is preserved in the resulting object. Second, when a defect approaches a boundary, fusion provides a systematic route to effective integrable boundary conditions and hence relates defect data to boundary data \cite{Bajnok:2007jg,Konechny:2015qla,Graham:2003nc}; more broadly, such questions are closely connected to the role of topological defects in renormalization group flows and, more recently, in non-invertible symmetries \cite{Chang:2018iay,Gaiotto:2012np}. Our main probe will be the defect $g$-function, which we compute from the defect TBA and compare between the fused configuration and the regime of large separation. In this way, we can track how the localized degrees of freedom are reorganized along fusion. We discuss topological and non-topological defects separately, and then turn to fusion with an integrable boundary.

\subsection{Topological defects}
\label{subsec:fusion_topological}

Topological defects are characterized by purely transmissive scattering and can be translated freely. In a massive IQFT, this property is reflected in the fact that the defect operator is diagonal in the bulk asymptotic basis. Correspondingly, the defect contributes an $\order{1}$ term to the free energy. For a single topological defect with transmission amplitude $T(\theta)$, the defect $g$-function is given by (cf.~Eq.~\eqref{eq:cal-gfunc})
\begin{equation}
\label{eq:logg_topo_single}
\log g_{d}(mR)=\int_{-\infty}^{\infty}\frac{\dd\theta}{2\pi}\,\Phi_{T}(\theta)\,\log\big(1+e^{-\varepsilon(\theta)}\big),
\qquad
\Phi_{T}(\theta):=-i\partial_{\theta}\log T(\theta),
\end{equation}
where $\varepsilon(\theta)$ is the bulk pseudo-energy determined solely by the bulk $S$-matrix, and therefore is independent of the presence of a purely transmitting defect.

\paragraph{Unfused defects.} 
Consider two topological defects $\mathcal{D}_{a}$ and $\mathcal{D}_{b}$ placed at different spatial positions. Since both defects are topological, in the closed channel their vacuum matrix element factorizes,
\begin{equation}
\label{eq:topo_factorization}
\expval{\mathcal{D}_{b}\mathcal{D}_{a}}=\expval{\mathcal{D}_{b}}\expval{\mathcal{D}_{a}}.
\end{equation}
Equivalently, in the open channel the non-extensive free energy is additive, implying
\begin{equation}
\label{eq:topo_logg_add}
\log g_{d,ab}(mR)=\log g_{d,a}(mR)+\log g_{d,b}(mR)
\end{equation}
for the two-defect system. At the level of scattering data this is consistent with the fact that sequential transmission through two defects multiplies the transmission factors,
\begin{equation}
\label{eq:T_product}
T_{ab}(\theta)=T_{a}(\theta)\,T_{b}(\theta)
\quad\Rightarrow\quad
\Phi_{T_{ab}}(\theta)=\Phi_{T_{a}}(\theta)+\Phi_{T_{b}}(\theta),
\end{equation}
which inserted into Eq.~\eqref{eq:logg_topo_single} reproduces Eq.~\eqref{eq:topo_logg_add}.

\paragraph{Fusion.} 
We define the fusion of two defects as the short-distance limit in which the separation between them is sent to zero and the pair is replaced by a single effective defect $\mathcal{D}_{a\star b}$. For topological defects this limit is smooth: because the defects can be moved freely, bringing them together does not generate additional singularities. The fused object is therefore again topological. In particular, in the IR limit $mR\to\infty$, the theory should reduce to a trivial topological field theory, so $\log g_{d}\to 0$ for each purely transmitting defect; hence the fused defect also has a trivial IR $g$-function.

Equation~\eqref{eq:same_defect_fusion} shows that the product of two topological defect operators of the same type, but with shifted defect parameters, splits into two contributions: the identity channel and the composite channel $D_{s-1}(\psi_{\lambda,\tilde{\lambda}})D_{s+1}(\psi_{\lambda,\tilde{\lambda}})$. In the closed-channel formulation it therefore gives a direct relation among unrenormalized defect one-point functions, while the finite defect $g$-function is obtained only after subtracting the corresponding defect free energies. Thus a direct-sum fusion rule should not be converted into a logarithmic $g$-function identity until the subtraction prescription.

Consequently, at the level of unrenormalized correlation functions the direct sum is represented by a plus sign. If we take $\mathcal{D}_{a\star b} \equiv D_{s}(\psi_{q\lambda,q^{-1}\tilde{\lambda}})\,D_{s}(\psi_{q^{-1}\lambda,q\tilde{\lambda}}) $, then
\begin{equation}
\begin{split}
\expval{\mathcal{D}_{a\star b}} = \expval{\rm id} + \expval{D_{s-1}(\psi_{\lambda,\tilde{\lambda}})D_{s+1}(\psi_{\lambda,\tilde{\lambda}}) } .
\end{split}
\end{equation}
\paragraph{Lee-Yang model}
For the Lee-Yang theory ${\cal M}(2,5) $, $q= \exp(2 i \pi /5)$ and this identity becomes
\begin{equation}
\expval{\mathcal{D}_{2\star 2}} = \expval{\rm id} + \expval{D_{2}(\psi_{\lambda,\tilde{\lambda}}) }.
\end{equation}
Here we have used the Kac label equivalence $(1,3)\sim(1,2)$. This relation is a statement about the unrenormalized fused defect operator. If one writes
\begin{equation}
\expval{\mathcal D_A}=e^{-f_A R}g_A(mR)
\end{equation}
for each superselection channel $A$, then direct-sum fusion gives
\begin{equation}
e^{-f_{2\star2}R}g_{2\star2}
=e^{-f_{\rm id}R}g_{\rm id}+e^{-f_2R}g_2 .
\end{equation}
Thus the finite $g$-function of the fused direct sum is not the logarithm of a single summand unless a dominant channel or a renormalized projection is specified. In the following comparison we therefore focus only on the finite, transmission-dependent part of the topological defect $g$-function after the defect free-energy subtraction. A complete operator-level treatment of the direct-sum subtraction remains to be supplied.

For the scaling Lee-Yang model, the transmission amplitudes for a topological defect are \cite{Bajnok:2007jg}
\begin{equation}
T_{-}(b)=[b+1][b-1], \quad T_{+}(b)=[5-b][-5-b],
\end{equation}
where 
\begin{equation}
[x]=i\frac{\sinh\left(\frac{\theta}{2}+i\frac{\pi x}{12}\right)}{\sinh\left(\frac{\theta}{2}+i\frac{\pi x}{12}-i\frac{\pi}{2}\right)}.
\end{equation}
The UV-IR correspondence \cite{Bajnok:2013waa} gives
\begin{equation}
\lambda = -\frac{|h_c|m^{6/5}}{2}e^{-i(b-2)\pi/5}, \quad \tilde{\lambda} = -\frac{|h_c|m^{6/5}}{2}e^{i(b-2)\pi/5},
\end{equation}
with
\begin{equation}
h_c = -\frac{\pi^{3/5}2^{4/5}5^{1/4}\sin\frac{2\pi}{5}}{\left(\Gamma(\frac{3}{5})\Gamma(\frac{4}{5})\right)^{1/2}}\left(\frac{\Gamma(\frac{2}{3})}{\Gamma(\frac{1}{6})}\right)^{6/5}.
\end{equation}
This implies that $D_{2}(\psi_{\lambda,\tilde{\lambda}})$ corresponds to the defect with $T_{-}(b)$, whereas $D_{2}(\psi_{q\lambda,q^{-1}\tilde{\lambda}})$ and $D_{2}(\psi_{q^{-1}\lambda,q\tilde{\lambda}})$ represent the defects with $T_{-}(b-2)$ and $T_{-}(b+2)$, respectively. The associated $g$-functions can be computed independently from Eq.~\eqref{eq:logg_topo_single}. We find that the transmission amplitudes satisfy
\begin{equation}\label{eq:trans_fusion}
T_{\pm}(b) = T_{\pm}(b-2) T_{\pm}(b+2) ,
\end{equation}
and the TBA equation only depends on the bulk $S$-matrix. As a result, the $g$-function contribution obeys
\begin{equation}
\log g_{2}(q\lambda,q^{-1}\tilde{\lambda}) + \log g_{2}(q^{-1}\lambda,q\tilde{\lambda}) = \log g_{2}(\lambda,\tilde{\lambda}).
\end{equation}
Equivalently, in the $b$-parametrization,
\begin{equation}
\Delta_{\rm finite}\log g_{d}
=\log g_{2}(b)-\log g_{2}(b-2)-\log g_{2}(b+2)=0.
\end{equation}
Thus, from the closed-channel point of view, the parameter-dependent finite part of the defect $g$-function cancels. The remaining channel-dependent information belongs to the defect free-energy subtraction rather than to the finite $g$-function itself. For the single defect, the defect free energy quoted in \cite{Bajnok:2013waa} is
\begin{equation}
f_{2}(b)=m\sin\left(\frac{b\pi}{6}\right),
\end{equation}
with analogous channel-dependent quantities for the fused direct sum.

The open-channel description gives a complementary perspective. If the fused pair can be represented by a single effective topological defect in the transmission sector, then the product formula in Eq.~\eqref{eq:trans_fusion} suggests that two defects with transmission amplitudes $T_{-}(b-2)$ and $T_{-}(b+2)$ are described effectively by the BAE. Similar fused-defect scattering ideas appear in integrable sine-Gordon defects \cite{Corrigan:2010aa}.
\begin{equation}
e^{ i L m \sinh \theta_i } T_{-}(\theta_{i}, b) \prod_{j=1}^{N} S_{i \mu_{j}}(\theta_{i\mu_{j}}) =1.
\end{equation}
This is the same BAE as that of a single topological defect with parameter $b$. From this viewpoint, the finite transmission-dependent defect $g$-function coincides with that of the single-defect description, giving
\begin{equation}
\log g_d^{\text{diff}}=0.
\end{equation}
Combining the closed- and open-channel viewpoints therefore suggests that the finite defect $g$-function is unchanged by topological fusion after the appropriate defect free-energy subtraction. At present, however, this should be regarded as an argument rather than a complete proof: an explicit defect TBA description of how the direct-sum defect operator ${\cal D}_{a\star b}$ acts on asymptotic states is still needed.

\subsection{Non-topological defects}
\label{subsec:fusion_nontopo}

We now turn to the fusion of non-topological integrable defects. The key qualitative difference from the topological case is the coexistence of reflection and transmission. As a consequence, the BAE requires a non-trivial diagonalization and already for a single defect one obtains two eigenmodes (the $\pm$ branches), see the discussion around Eq.~\eqref{eq:isingTR}.

\paragraph{Unfused defects.}
{Consider two defects $D_\alpha$ and $D_\beta$ with dimensionless separation $ma$, with $D_\alpha$ fixed at the origin. The unfused configuration is described by the state in Eq.~\eqref{eq:ZF_unfuse_state}. The effective amplitudes $\mathcal{R}_{\alpha\beta}(\theta,ma)$, $\tilde{\mathcal{R}}_{\alpha\beta}(\theta,ma)$ and $\mathcal{T}_{\alpha\beta}(\theta,ma)$ are given in Eqs.~\eqref{eq:ising_eff_R}--\eqref{eq:ising_eff_Rt}, and their diagonalization yields the BAEs in Eq.~\eqref{eq:BAE_two_nontopo}. For two identical defects $\alpha=\beta$, this reduces to Eq.~\eqref{eq:BAE_same_nontopo}. In the unfused configuration the reflected contribution carries an additional phase factor $e^{i ma\sinh\theta}$, which is absent in the fused limit. This phase is the origin of the strong oscillations observed in the integrand of unfused $g$-function.}

\paragraph{Fusion.}
{The fusion of the two defects is defined as the short-distance limit $ma\to 0$, in which the two separated defects are replaced by a single composite object. The corresponding fused state is}
\begin{equation}
A_{k,\alpha,\beta }^{\mu_{1} \ldots \mu_{N}}:=A_{\mu_{1}}\left(\theta_{\mu_{1}}\right) \ldots A_{\mu_{k}}\left(\theta_{\mu_{k}}\right) D_{\alpha} D_{\beta} \ldots A_{\mu_{N}}\left(\theta_{\mu_{N}}\right).
\end{equation}
At the level of scattering data, fusion amounts to evaluating the effective amplitudes at {$ma=0$}, and the BAEs reduce to the {$ma\to 0$} specialization of Eqs.~\eqref{eq:BAE_two_nontopo} and \eqref{eq:BAE_same_nontopo}.

\begin{figure}[tbp]
    \centering
    \includegraphics[width=0.55\linewidth]{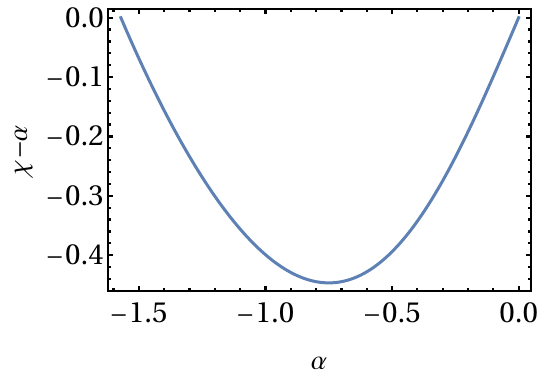}
    \caption{Shift of the effective defect parameter under fusion, shown as the difference between the fused parameter and the original one.}
    \label{fig:fusion_defect_param}
\end{figure}

{For the parity-invariant family in Eq.~\eqref{eq:isingTR}, the fused object can be reparametrized as a single defect with an effective parameter $\chi$. Here $\alpha$ and $\beta$ denote the parameters $\chi$ of the two original defects, so that their amplitudes are $T(\theta,\alpha),R(\theta,\alpha)$ and $T(\theta,\beta),R(\theta,\beta)$. Setting $ma=0$ in Eqs.~\eqref{eq:ising_eff_R}--\eqref{eq:ising_eff_T} gives}
\begin{equation}
\mathcal{T}_{\alpha\beta}^{(0)}(\theta)
=\frac{T(\theta,\alpha)T(\theta,\beta)}{1-R(\theta,\alpha)R(\theta,\beta)},\quad
\mathcal{R}_{\alpha\beta}^{(0)}(\theta)
=R(\theta,\alpha)+\frac{T(\theta,\alpha)^2R(\theta,\beta)}{1-R(\theta,\alpha)R(\theta,\beta)}.
\end{equation}
Substituting Eq.~\eqref{eq:isingTR}, one finds
\begin{equation}
\mathcal{T}_{\alpha\beta}^{(0)}(\theta)=T(\theta,\chi),\quad
\mathcal{R}_{\alpha\beta}^{(0)}(\theta)=R(\theta,\chi),
\end{equation}
provided the effective parameter is chosen as
\begin{equation}
\tan \chi=\frac{\sin\alpha+\sin\beta}{\cos\alpha\,\cos\beta}\,.
\end{equation}
Using the coupling parametrization in Eq.~\eqref{eq:isingTR}, the same relation can be written in terms of the coupling constants $g_{{\rm I},1}$, $g_{{\rm I},2}$ and the fused coupling $g_{\rm I}$ as
\begin{equation}
\label{eq:nontopo_g_fusion}
g_{\rm I}=\frac{g_{{\rm I},1}+g_{{\rm I},2}}{1+g_{{\rm I},1} g_{{\rm I},2}/4}\,.
\end{equation}
This matches the result in \cite{Delfino:1994nr} and is closely related to the conformal Ising defect fusion law studied in \cite{Bachas:2013nxa}. For two identical defects, the change of the effective parameter under fusion is maximal at
\begin{equation}
\alpha = -2 \tan ^{-1}\left(\sqrt{\frac{2}{\sqrt{3}}-1}\right),
\end{equation}
as shown in Fig.~\ref{fig:fusion_defect_param} within the trigonometric parametrization given above.

{In the defect TBA, the defect factor $P(\theta)$ entering the general Bethe-Yang equation \eqref{eq:gen_BAE} is precisely the $\pm$ eigenvalue obtained from the diagonalized BAEs, namely Eq.~\eqref{eq:BAE_two_nontopo} (and Eq.~\eqref{eq:BAE_same_nontopo} for identical defects). For the identical-defect configuration it is useful to display the corresponding eigenvalue and the integrand of the defect $g$-function explicitly as functions of the dimensionless separation $ma$:}
\begin{equation}
\label{eq:unfused_P_fpm}
\begin{split}
P_{\pm}(\theta,ma)&=\mathcal{T}_{\alpha\alpha}(\theta,ma)
\pm e^{i ma\sinh\theta}\mathcal{R}_{\alpha\alpha}(\theta,ma),\\
f^{\pm}(\theta,ma)&=
-i\,\frac{\partial}{\partial\theta}\log P_{\pm}(\theta,ma)\,
\log\left(1+e^{-mR\cosh\theta}\right),\\
\log g_d^{\pm}(ma)&=\int_{-\infty}^{\infty}\frac{\dd\theta}{2\pi}\,f^{\pm}(\theta,ma).
\end{split}
\end{equation}
{The additional integrand figures introduced below display $\operatorname{Re} (f^{\pm}(\theta,ma))$. Since the TBA weight is smooth on the real rapidity axis, sharp changes in these integrands are controlled by the logarithmic derivative of $P_{\pm}$. The plots of $|\mathcal{T}_{\alpha\alpha}(\theta,ma)|^2$ are therefore used as a diagnostic for the pole and zero structure of the diagonal eigenvalues $P_{-}$, while the value of the $g$-function is obtained only after integrating $f^{\pm}$. In what follows, we focus on the fusion of identical defects.}

\begin{figure}[tbp]
    \centering
    \begin{subfigure}[b]{0.49\linewidth}
        \centering
        \includegraphics[width=\textwidth]{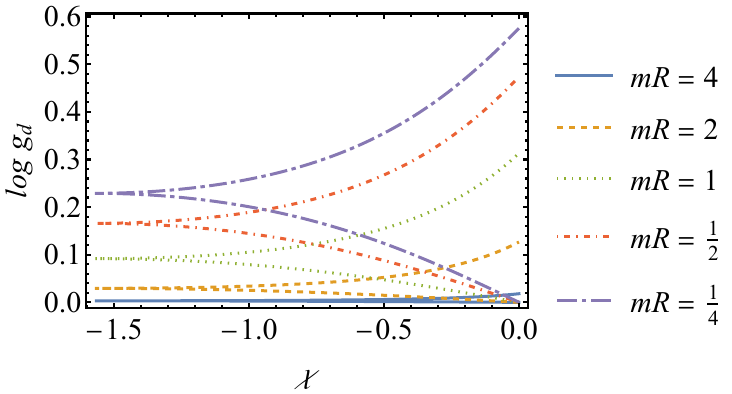}
        \caption{} 
        \label{fig:fused_chi}
    \end{subfigure}
    \hfill 
    \begin{subfigure}[b]{0.49\linewidth}
        \centering
        \includegraphics[width=\textwidth]{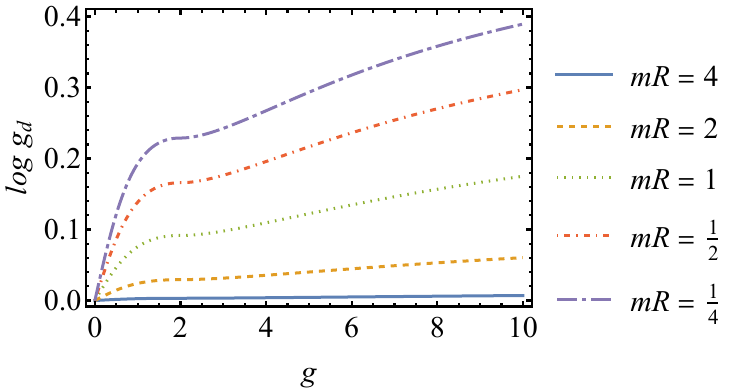}
        \caption{}
        \label{fig:fused_g}
    \end{subfigure}    
    \caption{Logarithmic defect $g$-function for fused identical defects, plotted as a function of the effective parameter $\chi$ (a) and of the Ising defect coupling $g_{\rm I}$ (b).}
\end{figure}

\emph{(i) Fused defects.}
{In the fused case ($ma=0$), the $g$-function is comparatively smooth near the boundary regime. Fusion enhances reflection, and already around $g_{\rm I}=2$ the system behaves almost as a boundary, as shown in Fig.~\ref{fig:fused_chi}. The fused $g$-function is monotonic in the parameter range shown, a behavior compatible with the expected direction of a genuine defect RG flow.}

\begin{figure}[tbp]
    \centering

    \begin{subfigure}[b]{0.6\linewidth} 
        \centering
        \includegraphics[width=\textwidth]{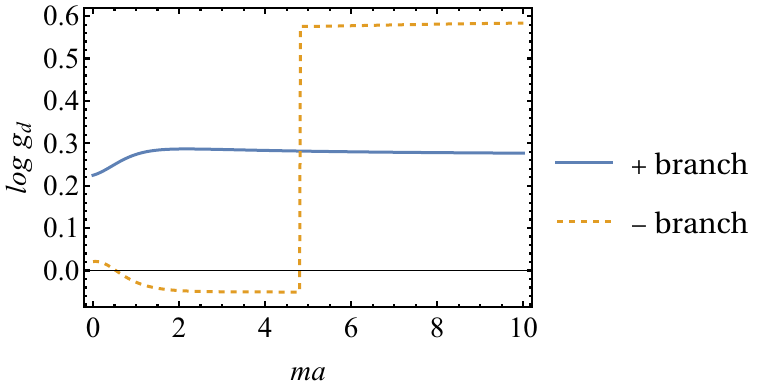}
        \caption{}
        \label{fig:unfuse_fix_chi}
    \end{subfigure}

    \vspace{0.8em}

    \begin{subfigure}[b]{0.48\linewidth}
        \centering
        \includegraphics[width=\textwidth]{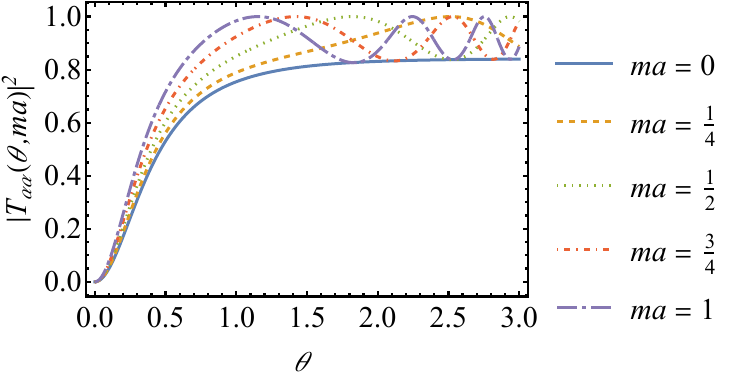}
        \caption{}
        \label{fig:unfuse_a_0_1}
    \end{subfigure}
    \hfill
    \begin{subfigure}[b]{0.48\linewidth}
        \centering
        \includegraphics[width=\textwidth]{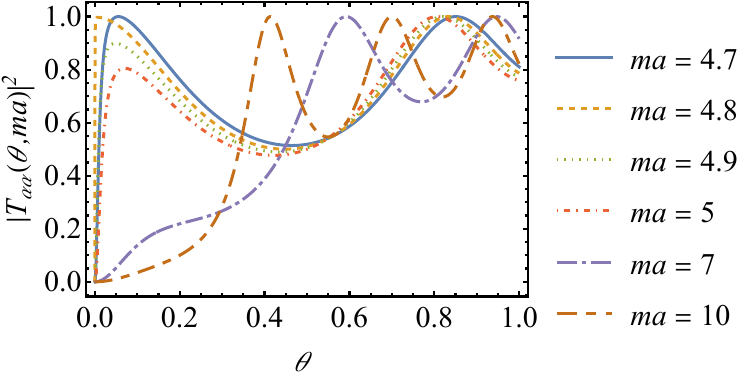}
        \caption{}
        \label{fig:unfuse_a_4_10}
    \end{subfigure}

    \caption{{Dependence on the dimensionless separation $ma$ for two identical unfused defects at fixed $\chi=-\frac{\pi}{15}$. (a) The logarithmic defect $g$-function as $ma$ varies. (b,c) The corresponding evolution of $|\mathcal{T}_{\alpha\alpha}(\theta,ma)|^2$ with $ma$.}}

\end{figure}

\emph{(ii) Unfused defects: discontinuity.}
{As the dimensionless separation $ma$ increases from zero, the phase $e^{i ma\sinh\theta}$ produces increasingly rapid oscillations in the effective amplitudes, as shown in Fig.~\ref{fig:unfuse_a_0_1}. Numerically, the first oscillatory peak of $|\mathcal{T}_{\alpha\alpha}(\theta,ma)|^2$ moves toward smaller rapidities. Since the TBA weight $\log(1+e^{-mR\cosh\theta})$ is largest near $\theta=0$, this migration enhances the low-energy part of the integral and can drive the $-$ branch of $\log g_d$ from positive values at small $ma$ to negative values at intermediate $ma$, as shown in Fig.~\ref{fig:unfuse_fix_chi}. We take $\chi=-\frac{\pi}{15}$ as a representative value; the same mechanism persists for small $|\chi|$.}

\begin{figure}[tbp]
    \centering

    \begin{subfigure}[b]{0.48\linewidth}
        \centering
        \includegraphics[width=\textwidth]{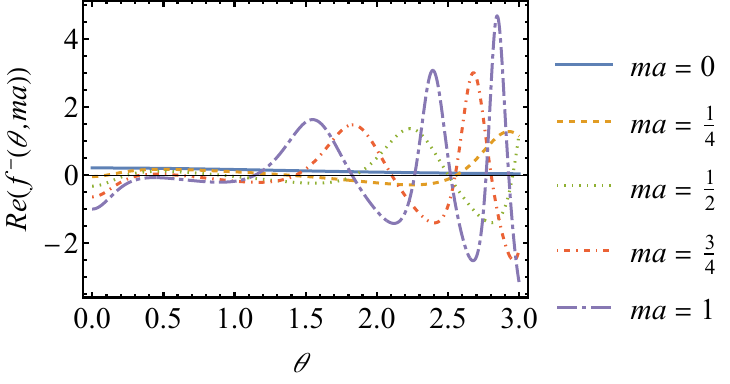}
        \caption{{$\operatorname{Re} (f^{-}(\theta,ma))$, $ma=0,\frac14,\frac12,\frac34,1$}}
    \end{subfigure}
    \hfill
    \begin{subfigure}[b]{0.48\linewidth}
        \centering
        \includegraphics[width=\textwidth]{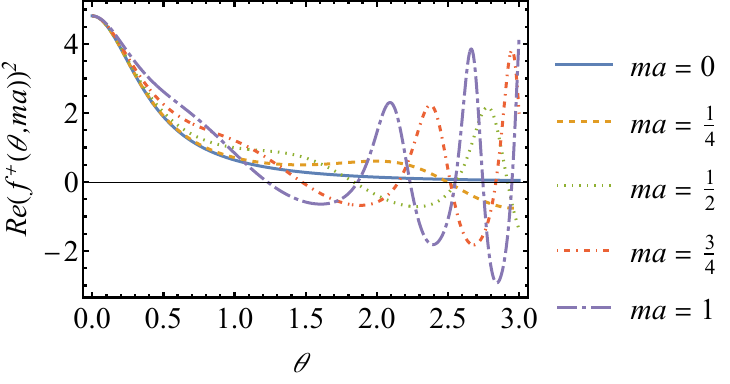}
        \caption{{$\operatorname{Re} (f^{+}(\theta,ma))$, $ma=0,\frac14,\frac12,\frac34,1$}}
    \end{subfigure}

    \vspace{0.8em}

    \begin{subfigure}[b]{0.48\linewidth}
        \centering
        \includegraphics[width=\textwidth]{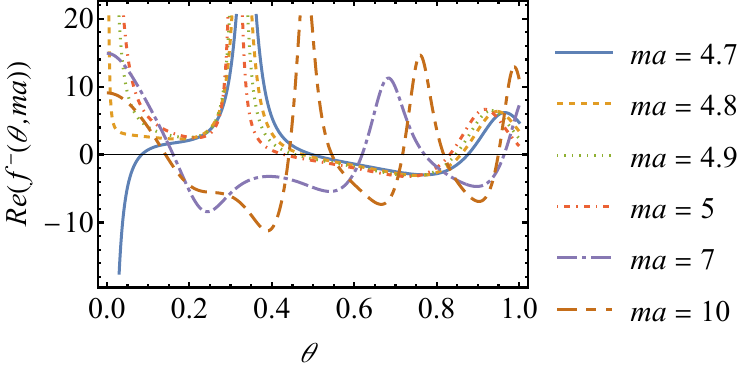}
        \caption{{$\operatorname{Re} (f^{-}(\theta,ma))$, $ma=4.7,4.8,4.9,5,7,10$}}
    \end{subfigure}
    \hfill
    \begin{subfigure}[b]{0.48\linewidth}
        \centering
        \includegraphics[width=\textwidth]{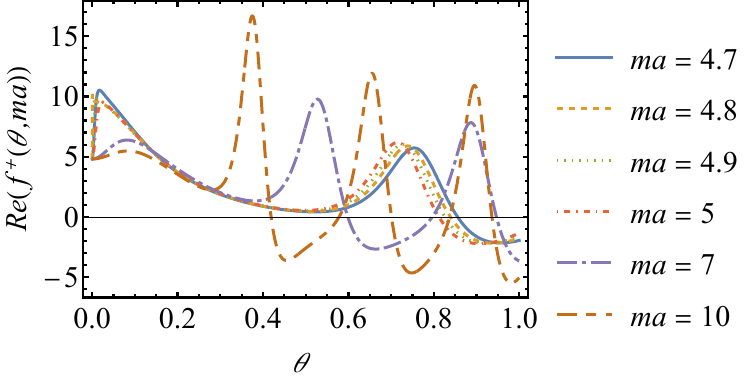}
        \caption{{$\operatorname{Re} (f^{+}(\theta,ma))$, $ma=4.7,4.8,4.9,5,7,10$}}
    \end{subfigure}

    \caption{{Comparison of $\operatorname{Re} (f^{\pm}(\theta,ma))$ for the same parameter choices as in Fig.~\ref{fig:unfuse_fix_chi}. The $-$ branch tracks the migration of the dominant oscillatory peak, while the $+$ branch remains smoother near $\theta=0$.}}
    \label{fig:unfuse_integrands_pm}
\end{figure}

{This interpretation is corroborated by the real part of the integrands in Eq.~\eqref{eq:unfused_P_fpm}. The plots of $\operatorname{Re} (f^{-}(\theta,ma))$ in Fig.~\ref{fig:unfuse_integrands_pm} show the same displacement of the dominant structure as Fig.~\ref{fig:unfuse_a_0_1} and Fig.~\ref{fig:unfuse_a_4_10}. In the first range, $ma=0,\frac14,\frac12,\frac34,1$, the peak is shifted continuously toward the low-rapidity region. In the second range, $ma=4.7,4.8,4.9,5,7,10$, the relevant singular structure reaches $\theta=0$, and the pole contribution from the logarithmic derivative enters the $g$-function. After this crossing the integral stabilizes again, in agreement with Fig.~\ref{fig:unfuse_fix_chi}.}

{The $+$ branch provides an important comparison. For $ma=0,\frac14,\frac12,\frac34,1$, the function $\operatorname{Re} (f^{+}(\theta,ma))$ remains relatively smooth near $\theta=0$, which accounts for the weak variation of the corresponding numerical curve. For $ma=4.7,4.8,4.9,5,7,10$, the oscillatory structure is still visible, but the pole at the origin is absent in $\operatorname{Re} (f^{+}(\theta,ma))$ within numerical accuracy. This suggests that, in the eigenvalue $P_+$, the transmission and reflection contributions cancel the singular part that survives in the $-$ branch. The stability of the $+$ branch is therefore a property of the diagonalized eigenvalue entering the TBA.}

\begin{figure}[tbp]
    \centering

    \begin{subfigure}[b]{0.48\linewidth} 
        \centering
        \includegraphics[width=\textwidth]{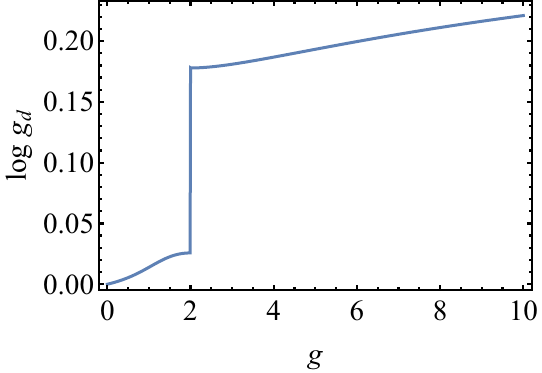}
        \caption{{$ma=\frac{1}{2}$}}
        \label{fig:unfuse_a_0.5}
    \end{subfigure}
    \hfil
    \begin{subfigure}[b]{0.48\linewidth}
        \centering
        \includegraphics[width=\textwidth]{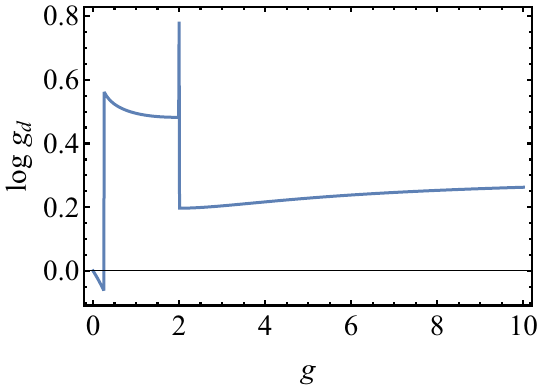}
        \caption{{$ma=4$}}
        \label{fig:unfuse_a_4}
    \end{subfigure}
    \hfill
    \caption{{Logarithmic defect $g$-function for two identical unfused defects as a function of the Ising defect coupling $g_{\rm I}$.}}
    \label{fig:unfuse_change_a}
\end{figure}

{If instead we fix $ma$ and vary $\chi$, the discontinuities persist, as shown in Fig.~\ref{fig:unfuse_change_a}. One of them is associated with the boundary limit $\chi=-\frac{\pi}{2}$, equivalently $g_{\rm I}=2$, where the defect becomes purely reflective; this limit is discussed in point \emph{(iii)}. The other appears at small $|\chi|$ for sufficiently large $ma$, again because the first oscillatory peak reaches $\theta=0$. The two branches therefore differ generically: the unfused effective defect retains an $ma$-dependent, bulk-like contribution, and this distinction is particularly pronounced on the $-$ branch.}

\begin{figure}[tbp]
    \centering

    \begin{subfigure}[b]{0.48\linewidth}
        \centering
        \includegraphics[width=\textwidth]{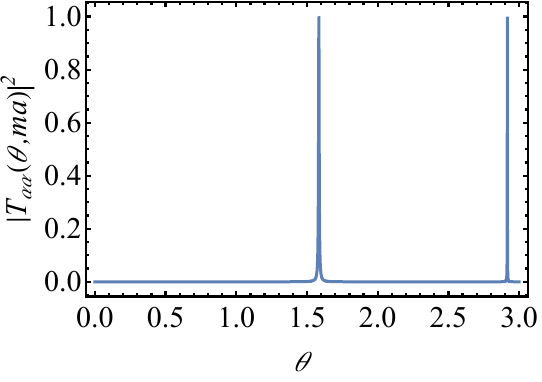}
        \caption{{$ma=\frac{1}{2}$}}
    \end{subfigure}
    \hfil
    \begin{subfigure}[b]{0.48\linewidth}
        \centering
        \includegraphics[width=\textwidth]{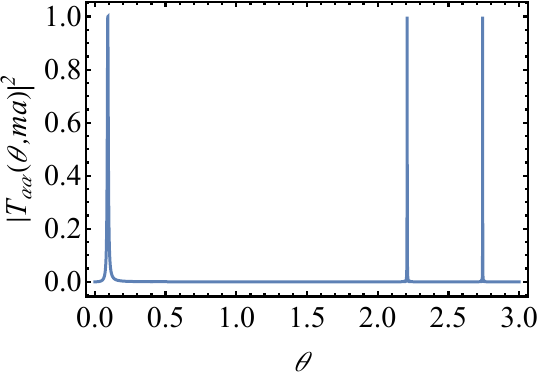}
        \caption{{$ma=1$}}
    \end{subfigure}

    \vspace{0.8em}

    \begin{subfigure}[b]{0.48\linewidth}
        \centering
        \includegraphics[width=\textwidth]{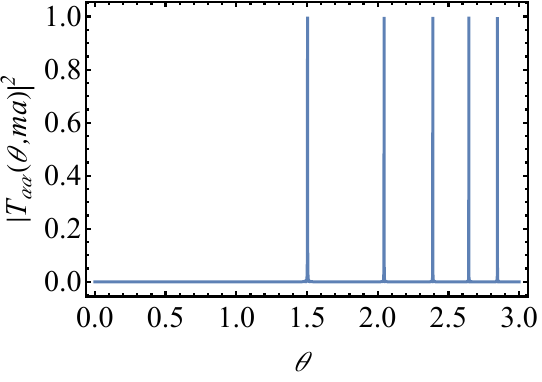}
        \caption{{$ma=2$}}
    \end{subfigure}
    \hfill
    \begin{subfigure}[b]{0.48\linewidth}
        \centering
        \includegraphics[width=\textwidth]{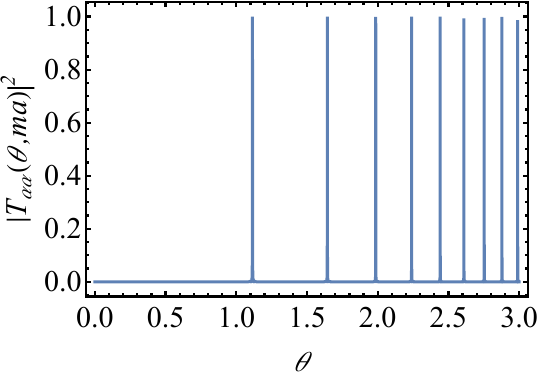}
        \caption{{$ma=3$}}
    \end{subfigure}

    \caption{{Evolution of $|\mathcal{T}_{\alpha\alpha}(\theta,ma)|^2$ with $ma$ at fixed $\chi=-\frac{\pi}{2.1}$.}}
    \label{fig:tsquare_close_boundary}
\end{figure}

\emph{(iii) Near-boundary regime and continuity at $g_{\rm I}=2$.}
{At $g_{\rm I}=2$, or equivalently $\chi=-\frac{\pi}{2}$, the transmission amplitude vanishes and the defect becomes purely reflective. For two separated defects with $ma$ not too small and $g_{\rm I}\simeq 2$, transmission is strongly suppressed, except for a set of narrow resonant peaks. As $ma$ increases, more such peaks appear, whereas they disappear in the strict boundary limit, as shown in Fig.~\ref{fig:tsquare_close_boundary}. Away from this limit, the resonances compensate the bulk-like contribution and thereby preserve the continuity of the full $g$-function. A natural interpretation is that they describe resonant transmission between the two defects: only specific modes can propagate in the finite segment, and once the boundary limit is reached these modes cease to exist.}

\begin{figure}[tbp]
    \centering

    \begin{subfigure}[b]{0.48\linewidth}
        \centering
        \includegraphics[width=\textwidth]{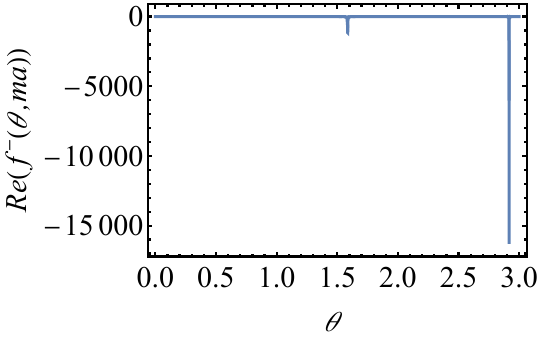}
        \caption{{$ma=\frac{1}{2}$}}
    \end{subfigure}
    \hfil
    \begin{subfigure}[b]{0.48\linewidth}
        \centering
        \includegraphics[width=\textwidth]{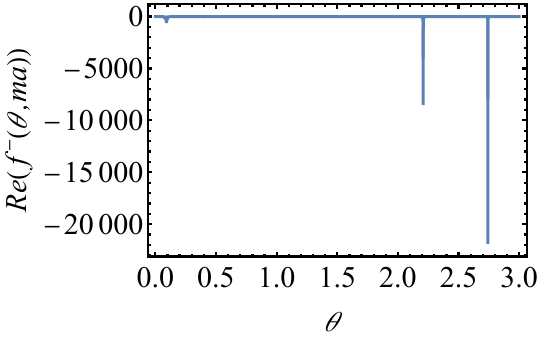}
        \caption{{$ma=1$}}
    \end{subfigure}

    \vspace{0.8em}

    \begin{subfigure}[b]{0.48\linewidth}
        \centering
        \includegraphics[width=\textwidth]{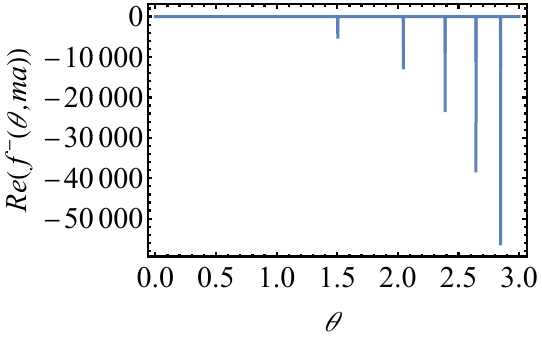}
        \caption{{$ma=2$}}
    \end{subfigure}
    \hfill
    \begin{subfigure}[b]{0.48\linewidth}
        \centering
        \includegraphics[width=\textwidth]{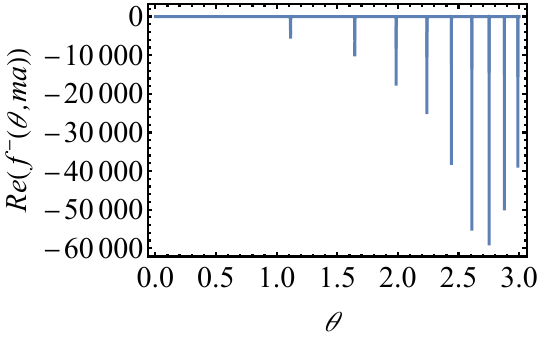}
        \caption{{$ma=3$}}
    \end{subfigure}

    \caption{{Near-boundary behavior of $\operatorname{Re} (f^{-}(\theta,ma))$ at fixed $\chi=-\frac{\pi}{2.1}$. The resonant structures occur in the same rapidity regions as the peaks of $|\mathcal{T}_{\alpha\alpha}(\theta,ma)|^2$ in Fig.~\ref{fig:tsquare_close_boundary}.}}
    \label{fig:fminus_close_boundary}
\end{figure}

\begin{figure}[tbp]
    \centering

    \begin{subfigure}[b]{0.48\linewidth}
        \centering
        \includegraphics[width=\textwidth]{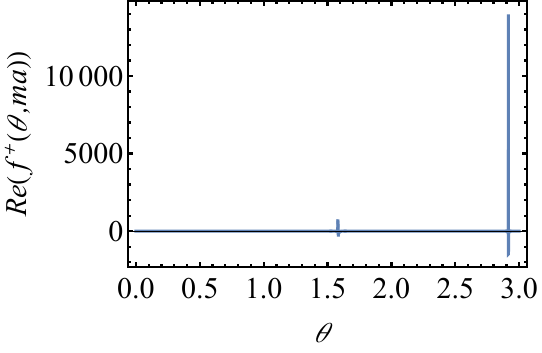}
        \caption{{$ma=\frac{1}{2}$}}
    \end{subfigure}
    \hfil
    \begin{subfigure}[b]{0.48\linewidth}
        \centering
        \includegraphics[width=\textwidth]{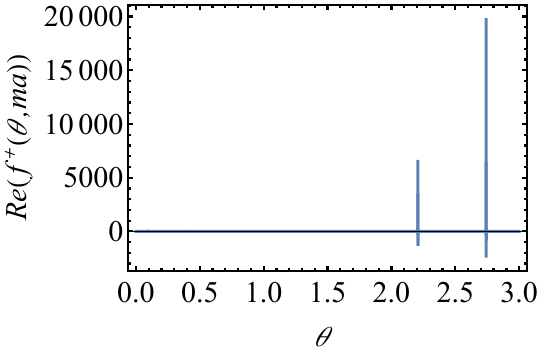}
        \caption{{$ma=1$}}
    \end{subfigure}

    \vspace{0.8em}

    \begin{subfigure}[b]{0.48\linewidth}
        \centering
        \includegraphics[width=\textwidth]{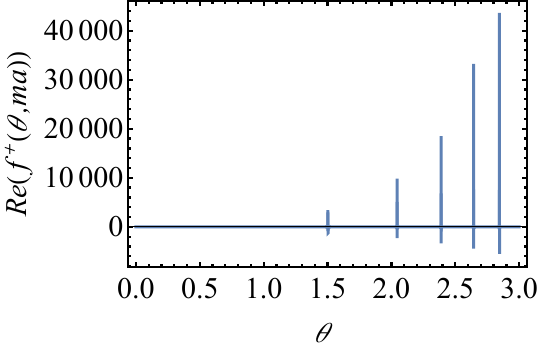}
        \caption{{$ma=2$}}
    \end{subfigure}
    \hfill
    \begin{subfigure}[b]{0.48\linewidth}
        \centering
        \includegraphics[width=\textwidth]{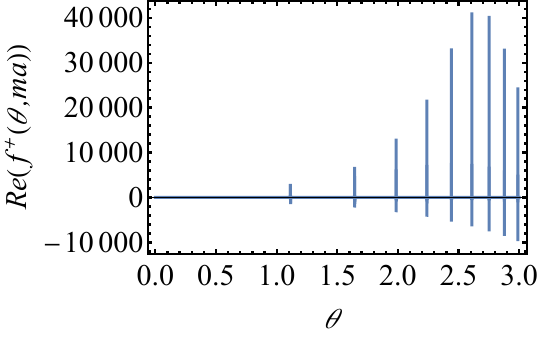}
        \caption{{$ma=3$}}
    \end{subfigure}

    \caption{{Near-boundary behavior of $\operatorname{Re} (f^{+}(\theta,ma))$ at fixed $\chi=-\frac{\pi}{2.1}$. The branch shows the same resonant mechanism as $f^{-}$, with branch-dependent numerical differences.}}
    \label{fig:fplus_close_boundary}
\end{figure}

{The same conclusion is visible at the level of $\operatorname{Re} (f^{\pm}(\theta,ma))$. Figures~\ref{fig:fminus_close_boundary} and \ref{fig:fplus_close_boundary} show that the narrow structures in the real parts of the two integrands occur in the same rapidity regions as the resonant peaks in $|\mathcal{T}_{\alpha\alpha}(\theta,ma)|^2$, up to branch-dependent normalization and sign. The smooth TBA weight cannot generate such narrow peaks by itself; it only biases their contribution toward small rapidity. Thus the behavior of $|\mathcal{T}_{\alpha\alpha}(\theta,ma)|^2$ provides a reliable guide to the singular part of the integrands, while the final value of $\log g_d$ is determined by the weighted integral of $f^{\pm}$.}

In the boundary case, the system acquires the finite-size bulk contribution
\begin{equation}
\label{eq:boundary_limit}
 ma \int_{-\infty}^{\infty} \frac{\dd\theta}{2\pi}\, \cosh\theta\,\log\big(1+e^{-mR\cosh\theta}\big),
\end{equation}
{which originates from the exponential factor $e^{i ma\sinh\theta}$ in the BAE. The coefficient shown here assumes the one-pass phase convention of Eq.~\eqref{eq:BAE_same_nontopo}; if the boundary-limit quantization is written with a two-pass phase, this term would have a factor of two. If this bulk term is not separated from the localized defect contribution, the limiting value appears discontinuous. After it is identified as the finite-size contribution of the segment between the two reflective defects, the remaining defect contribution is continuous across the approach to $g_{\rm I}=2$.}

\begin{figure}[tbp]
    \centering
    \includegraphics[width=0.55\linewidth]{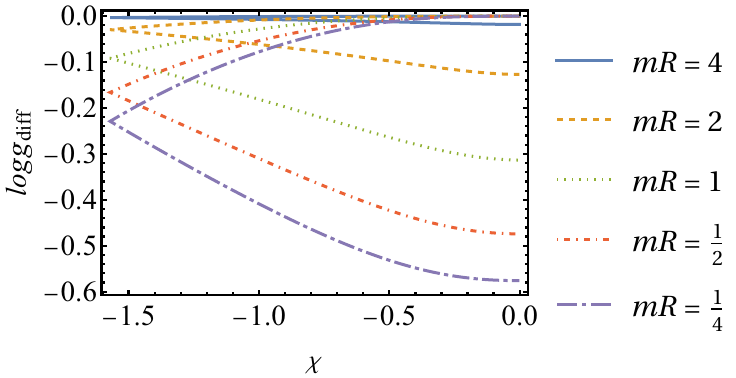}
    \caption{{Change in the logarithmic defect $g$-function during fusion as the defect parameter $\chi$ is varied.}}
    \label{fig:diff_decouple}
\end{figure}

\emph{(iv) Decoupling regime.}
{In the limit $ma\to\infty$, the two non-topological defects decouple, and the logarithmic $g$-function becomes additive, i.e. it is given by the sum of the individual defect contributions. As shown in Fig.~\ref{fig:diff_decouple}, it decreases in the non-topological Ising examples studied here and remains unchanged in the topological limit.}

{Finally, unlike the fused configuration, the unfused system depends on both the dimensionless separation $ma$ and the defect coupling, leading to a richer phenomenology. In particular, the decoupled limit is recovered as expected, while the finite $g$-function decreases or remains unchanged under fusion in the examples studied here. When one of the two defects is tuned to a boundary while the other remains a defect, the same formalism still applies and connects smoothly to defect-boundary fusion. However, this limit is too restrictive for our purposes, since the defect can reduce only to the fixed boundary condition. To study fusion with more general boundary conditions, we therefore turn to integrable boundaries explicitly.}

\subsection{Defect between boundaries}
\label{subsec:defect_boundary}

Having analyzed integrable defects and their fusion, we now turn to the fusion of a defect with an integrable boundary. This provides the massive scattering counterpart of the dCFT statement in Sec.~\ref{sec:ID}: when a topological defect approaches a boundary, defect fields reduce to boundary fields \cite{Runkel:2010ym}. From the scattering viewpoint, we therefore expect the fusion to be encoded in an effective boundary condition. In what follows we make this expectation explicit at the level of the BAE and the defect $g$-function.

\paragraph{Topological defect with boundary}
For a purely transmitting defect that breaks parity invariance and is inserted between two integrable boundaries, the ZF algebra yields
\begin{equation}
\begin{split}
e^{2 i L m \sinh \theta_i } K_{a}(\theta_i) T_{-}(\theta_i) K_{b}(\theta_i) T_{+}(\theta_i) \prod_{l=1}^{N} S_{i\mu_{l}}(\theta_{i}-\theta_{\mu_l}) S_{\mu_{l} i}(\theta_{i}+\theta_{\mu_l}) =1 .
\end{split}
\end{equation}
It follows that the fusion of the defect with the boundary factor $K_b$ is encoded in the effective reflection amplitude
\begin{equation}
T_{-}(\theta)\,K_{b}(\theta)\,T_{+}(\theta).
\end{equation}
This agrees with the result of Ref.~\cite{Bajnok:2007jg} and provides a scattering interpretation of the dCFT fusion rule: the original boundary state is fused to a new boundary state. In practice, this composition offers a systematic way to generate new integrable boundary conditions.

For the fused boundary, the $g$-function takes the form
\begin{equation}\label{eq:topo-bdry-gfunction}
\begin{split}
\log g_{d,b}^{\text{fused}} &= \int_{-\infty}^\infty \frac{\dd\theta}{4\pi} \left( -i \frac{\partial}{\partial \theta} \log \big[T_{-}(\theta) K_{b}(\theta) T_{+}(\theta)\big]\right) \log \left(1 + e^{- \varepsilon (\theta)}\right) + \text{Const},
\end{split}
\end{equation}
where the constant term is the ratio of Fredholm determinants and is determined solely by the bulk $S$-matrix. The pseudo-energy satisfies the standard boundary TBA equation
\begin{equation}\label{eq:TBA_bdry}
\begin{split}
\varepsilon (\theta ) =mR \cosh \theta-\int_{0}^{\infty} \frac{\dd \theta^\prime}{2 \pi } \big(\varphi\left(\theta -\theta^\prime\right)+ \varphi\left(\theta +\theta^\prime\right)\big)\log\left(1+ e^{-\varepsilon \left(\theta^\prime\right) }\right),
\end{split}
\end{equation}
which can be rewritten as
\begin{equation}
\varepsilon (\theta ) =mR \cosh \theta-\int_{-\infty}^{\infty} \frac{\dd \theta^\prime}{2 \pi } \varphi\left(\theta -\theta^\prime\right)\log\left(1+e^{-\varepsilon \left(\theta^\prime\right) }\right),
\end{equation}
by extending the pseudo-energy to negative rapidities via $\varepsilon(-\theta)=\varepsilon(\theta)$. In this form the boundary TBA has the same convolution structure as the periodic bulk TBA after even extension, although it still describes the boundary quantization problem.

Without fusion, the defect and boundary can be decoupled. Since the decoupled topological defect contributes through purely right- or left-moving scattering, the total logarithmic $g$-function becomes
\begin{equation}
\begin{split}
\log g_{d,b}^{\text{unfused}} &= \int_{-\infty}^\infty \frac{\dd\theta}{4\pi} \left( -i \frac{\partial}{\partial \theta} \log  K_{b}(\theta) \right) \log \left(1 + e^{- \varepsilon(\theta)}\right) + \text{Const}\\
& + \int_{-\infty}^\infty \frac{\dd\theta}{2\pi} \left( -i \frac{\partial}{\partial \theta} \log  T_{-}(\theta) \right) \log \left(1 + e^{- \varepsilon (\theta)}\right) .
\end{split}
\end{equation}
Using the topological-defect unitarity relation $T_+(\theta)T_-(-\theta)=1$ and the evenness of the TBA weight, the $T_+$ term in the fused expression can be rewritten as the second half of the full-line $T_-$ contribution. Consequently, the two transmission contributions in the fused expression reproduce the single full-line topological-defect contribution in the unfused expression. Therefore
\begin{equation}
\begin{split}
\log g_{\text{diff}}= \log g_{d,b}^{\text{fused}} - \log g_{d,b}^{\text{unfused}} = 0.
\end{split}
\end{equation}
In contrast to the fusion of non-topological defects, the value of $\log g_{d,b}$ remains invariant under this fusion process, indicating that the effective localized degrees of freedom are preserved.

\paragraph{Non-topological defect with boundary}
We next consider two integrable boundaries with a parity-invariant non-topological defect in the bulk. The defect transmission and reflection amplitudes are given in Eqs.~\eqref{eq:isingTR} and \eqref{eq:reflectiona}. For the Ising integrable boundary we use \cite{Ghoshal:1993tm,Castro-Alvaredo:2008fni}
\begin{equation}
K_{a}(\theta, \kappa_a)=i \tanh \left(-\frac{\theta }{2}+\frac{i \pi }{4}\right) \frac{ \kappa_a -i \sinh (\theta )}{\kappa_a +i \sinh (\theta )}.
\end{equation}
The limit $\kappa_a \to \infty$ corresponds to the fixed boundary condition, while $\kappa_a =1$ gives the free boundary condition. Placing the boundaries $K_a$ and $K_b$ at $y=-L$ and $y=0$, respectively, and the defect at $y=-a$, the ZF algebra yields
\begin{equation}
e^{2 i L m  \sinh \theta_i} K_{a}(\theta_i) \mathcal{R}_{\alpha, b}(\theta_i) \prod_{l=1}^{N} S_{i \mu_l} S_{\mu_l i} = 1,
\end{equation}
where
\begin{equation}
\begin{split}
\mathcal{R}_{\alpha,b}(\theta_i) &= e^{-2 i m a \sinh \theta_i} R_{\alpha}(\theta_i, -a)  +  \frac{ T_{\alpha}(\theta_i)^2 K_{b}(\theta_i)}{ 1- \tilde{R}_{\alpha}(\theta_i, -a) K_{b}(\theta_i)} \\
& = R_{\alpha}(\theta_i, 0)  +  \frac{ T_{\alpha}(\theta_i)^2 K_{b}(\theta_i)}{ 1- \tilde{R}_{\alpha}(\theta_i, -a) K_{b}(\theta_i)}.
\end{split}
\end{equation}
Thus, in left-to-right scattering the defect together with the boundary $K_b$ forms an effective boundary, and the contribution of multiple reflections between them is essential.

{The discussion below is organized as a comparison between the fused and separated descriptions: the fused effective boundary, its limiting checks, the fused $g$-function, the finite-separation configuration, and the resulting change under fusion.}

{\emph{(i) Fused effective boundary.}} {In the strict fusion limit $ma=0$, the defect and boundary are replaced by a single effective boundary. In this paragraph $\gamma$ denotes the parameter of the non-topological Ising defect in Eq.~\eqref{eq:isingTR}, so that the defect amplitudes are $T(\theta,\gamma)$ and $R(\theta,\gamma)$. The effective reflection factor in the fusion limit is obtained from the multiple-reflection sum above by setting $a=0$:}
\begin{equation}
\mathcal{R}_{\gamma,b}^{(0)}(\theta)
=R(\theta,\gamma)+\frac{T(\theta,\gamma)^2 K_b(\theta,\kappa_b)}
{1-R(\theta,\gamma)K_b(\theta,\kappa_b)}.
\end{equation}
{Substitution of Eq.~\eqref{eq:isingTR} and the boundary reflection factor $K_b(\theta,\kappa_b)$ shows that this effective amplitude has the same functional form as a single Ising boundary reflection factor $K(\theta,\kappa)$,}
\begin{equation}
\mathcal{R}_{\gamma,b}^{(0)}(\theta)=K(\theta,\kappa_f),
\end{equation}
provided the fused boundary parameter is
\begin{equation}
\kappa_f = \frac{ \kappa_b (1-\sin \gamma )+2 \sin \gamma }{\sin \gamma +1}.
\end{equation}
{Thus the fusion changes only the boundary parameter, not the functional form of the Ising reflection factor. For $\kappa_b=1$ and $\kappa_b\to\infty$, the fused parameter corresponds to the same boundary condition. For $1<\kappa_b<\infty$, the defect drives the boundary toward the fixed condition. In the separated problem, the same tendency becomes more pronounced when $\kappa_b$ is itself closer to the free boundary.}

{\emph{(ii) Limiting checks.}} {The two special values of $\gamma$ provide checks of this effective-boundary interpretation. At $\gamma= -\pi /2$ the BAE reduces to}
\begin{equation}
e^{2 i m (L-a) \sinh \theta_i} K_{a}(\theta_i) R_{\alpha}(\theta_i, -a) \prod_{l=1}^{N} S_{i \mu_l} S_{\mu_l i} = 1,
\end{equation}
{which shows that the two subsystems decouple: the particle cannot penetrate into the segment of length $a$ on the other side.}

{At $\gamma=0$, by contrast, the defect reflection vanishes and only transmission remains, reducing the system to the case of two boundaries with a topological defect. The corresponding parameter mapping implies that the fused boundary parameter coincides with the unfused one, since in the Ising model the topological defect has unit transmission.}

\begin{figure}[tbp]
    \centering
    \includegraphics[width=0.55\linewidth]{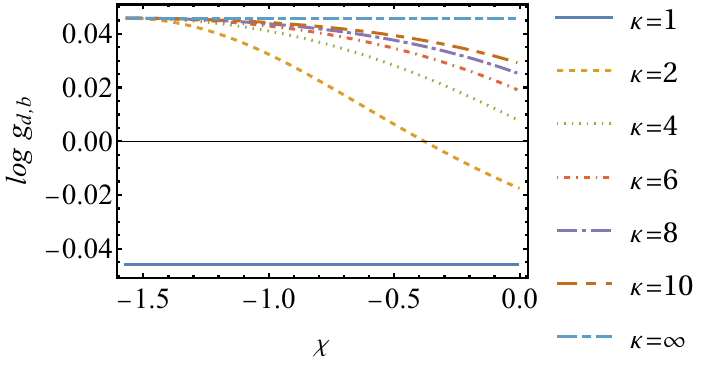}
    \caption{Logarithmic $g$-function of the fused boundary as a function of the parameter $\chi$.}
    \label{fig:fusion_defect_bdry_chi}
\end{figure}

{\emph{(iii) Fused $g$-function contribution.}} {Once the effective reflection factor is fixed, the joint contribution of boundary $K_b$ and the non-topological defect to the $g$-function is}
\begin{equation}\label{eq:generalg_nontopo_bdry}
\log g_{d,b} = \int_{-\infty}^\infty \frac{\dd \theta}{4\pi} \left( -i \frac{\partial}{\partial \theta} \log \mathcal{R}_{\alpha, b} (\theta) \right) \log \left(1 + e^{-m R \cosh \theta}\right).
\end{equation}
{The integral is taken over the full real line because the defect renders the imaginary part symmetric about the origin and the real part symmetric about $\theta=0$. For $ma=0$, this expression gives the parameter-dependent part of the fused-boundary $g$-function and is consistent with the boundary-parameter composition discussed above, as illustrated in Fig.~\ref{fig:fusion_defect_bdry_chi}. The Fredholm-determinant term is independent of the defect-boundary parameter in the comparisons below and cancels from the plotted differences.}

{\emph{(iv) Finite-separation configuration.}} {The comparison problem is the unfused configuration, in which the defect and the boundary are separated by a distance $a$. This configuration depends on three independent parameters and therefore exhibits a considerably richer behavior, including discontinuities. At the same time, the underlying mechanism is essentially the same as for unfused non-topological defects: it is again controlled by the structure of $|\mathcal{R}_{\alpha,b}(\theta)|^{2}$.}

{Since our main interest here is the change of the $g$-function between the decoupled and fused configurations, and since the full three-parameter regime is numerically much less stable, we will not attempt a systematic analysis of the unfused case. The qualitative features can nevertheless be understood from the profile of $|\mathcal{R}_{\alpha,b}(\theta)|^{2}$: for instance, as the oscillations approach the $\theta$ axis and the peak moves toward $\theta=0$, the profile may invert in a manner reminiscent of the fixed-boundary reflection pattern.}

{\emph{(v) Fusion-induced change in the $g$-function.}} {The final step is to compare the decoupled and fused configurations. As in the topological case, if we decouple the non-topological defect from the boundary, the logarithmic $g$-function becomes additive}
\begin{equation}
\begin{split}
\log g_{d,b}^{\text{unfused}} &= \int_{-\infty}^\infty \frac{\dd \theta}{2\pi} \left( -i \frac{\partial}{\partial \theta} \log \big( T_{\alpha}(\theta) \pm  R_{\alpha}(\theta)  \big) \right) \log \left(1 + e^{-m R \cosh \theta}\right)\\
& + \int_{-\infty}^\infty \frac{\dd \theta}{4\pi}\left( -i \frac{\partial}{\partial \theta} \log K_{b}(\theta) \right) \log \left(1 + e^{-m R \cosh \theta}\right) 
\end{split}
\end{equation}
{whereas the fused result is simply Eq.~\eqref{eq:generalg_nontopo_bdry} evaluated at $ma=0$. For the special cases $\kappa_b =1$ or $\kappa_b \to \infty$, the difference can be written as}\begin{equation}
\log g^{\text{diff}} = - \int_{-\infty}^\infty \frac{\dd \theta}{2\pi} \left( -i \frac{\partial}{\partial \theta} \log \big( T_{\alpha}(\theta) \pm  R_{\alpha}(\theta)  \big) \right) \log \left(1 + e^{-m R \cosh \theta}\right),
\end{equation}
which is generally negative in the parameter range studied here and vanishes in the topological case. This implies that, during the fusion of a non-topological defect with either a fixed or a free boundary, the finite localized contribution to the $g$-function decreases.

\begin{figure}[tbp]
    \centering
    \includegraphics[width=0.55\linewidth]{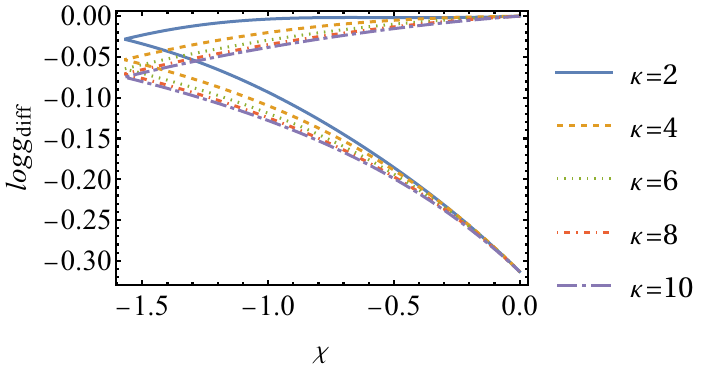}
    \caption{Change in the logarithmic $g$-function under defect-boundary fusion as a function of $\chi$.}
    \label{fig:fusion_defect_bdry_diff}
\end{figure}

{For a general boundary, with $mR =1$, Fig.~\ref{fig:fusion_defect_bdry_diff} shows that throughout the parameter space studied here the $g$-function is reduced after fusion, indicating a decrease in the localized degrees of freedom. Together with the non-topological defect-fusion examples above, this suggests the conjecture that fusion processes involving non-topological defects decrease the finite localized $g$-function contribution.}

\section{Conclusions and outlook}
\label{sec:conclude}

In this work, we developed a unified framework for the fusion of integrable defects based on the exact defect $g$-function. Starting from the Bethe-ansatz quantization conditions derived within the Zamolodchikov-Faddeev algebra, we formulated a defect TBA description of the exact $g$-function and used it as a quantitative probe of fusion. This framework provides a common language that relates the scattering data, finite-size thermodynamics, and the effective localized degrees of freedom carried by defects and boundaries.

\paragraph{Summary of the main results.}
For topological defects, the vacuum matrix elements factorize in the separated configuration and the logarithmic $g$-function is additive. The fusion limit is expected to be smooth from the topological nature of the defect, and the transmission factors combine multiplicatively in agreement with the additive structure of $\log g_d$. In the scaling Lee-Yang theory, the transmission amplitude together with the UV-IR map supports the finite defect $g$-function relation after the appropriate defect free-energy subtraction. Any remaining channel-dependent information belongs to the defect free-energy subtraction rather than to the finite $g$-function itself.

For non-topological defects, the coexistence of reflection and transmission breaks the topological property and leads to qualitatively richer behavior. We derived the effective amplitudes for two separated defects and obtained the corresponding quantization conditions. In the fused limit, the effective amplitudes reduce to those of a single composite defect, and in the parity-invariant Ising model this yields an exact composition law for the defect parameter. The fused $g$-function is correspondingly much smoother and is monotonic in the parameter range studied near the boundary regime.

By contrast, the unfused non-topological configuration exhibits genuinely separation-dependent effects. The phase $e^{i ma\sinh\theta}$ generates strong oscillations in the effective amplitudes and may induce discontinuities in $\log g_d$ when poles approach the $\theta=0$ axis. Near the boundary regime, however, resonant transmission remains essential for preserving the continuity of the full $g$-function despite sharp changes in the integrand. In the decoupling limit $ma\to\infty$, the logarithmic $g$-function becomes additive again. In the examples studied here, the finite localized contribution decreases under non-topological fusion, while it is unchanged in the topological limit.

We also studied defect-boundary fusion. For topological defects, fusion with a boundary is encoded by the effective reflection amplitude $T_{-}K_{b}T_{+}$, which provides a scattering realization of the dCFT statement that defect perturbations reduce to boundary perturbations as the defect approaches the boundary. In this case the finite logarithmic $g$-function remains unchanged under fusion. For non-topological defects, by contrast, multiple reflections between the defect and the boundary are essential and generate a nontrivial boundary parameter, with a simple analytic composition law in the fused Ising model. The corresponding shift in the examples studied here is negative, indicating a reduction in localized degrees of freedom under fusion with a boundary.

Taken together, these results establish the defect $g$-function as a sensitive quantitative diagnostic of integrable defects and their fusion, and make explicit the sharp distinction between the topological and non-topological cases.

\paragraph{Discussion and outlook.}
Several open questions remain. First, it would be important to formulate the defect TBA for the fused topological defect, both at the level of the BAE and at the operator level. Such a formulation could provide quantitative control over the change of the $g$-function under fusion and help clarify the relation between scattering fusion and categorical fusion in perturbed CFT.

Second, it would be desirable to extend our diagonalization method to general non-diagonal cases. Solving the non-diagonal problem would make it possible to compute defect $g$-functions in a wider class of defects and would therefore broaden the study of their RG flows.

More generally, it would be particularly interesting to establish whether there is a model-independent monotonicity property for the finite defect $g$-function under fusion, and to determine precisely when fusion can be interpreted as an RG flow.

\acknowledgments
We are indebted to Yi-Jun He, Shan-Ping Wu and Jing-Cheng Chang for very helpful discussions. The work of Y.J. is supported by Startup Fund No.4007022326 of Southeast University and National Natural Science Foundation of China through Grant No.12575073 and 12247103. The work of Yu-Xiao Liu is supported by the National Natural Science Foundation of China (Grant No. 12247101), the Fundamental Research Funds for the Central Universities (Grant No. lzujbky-2025-jdzx07), the Natural Science Foundation of Gansu Province (No. 22JR5RA389, No.25JRRA799), and the '111 Center' under Grant No. B20063.

\appendix

\section{Details for defect Bethe ansatz equations}
\label{app:reduce}

In this appendix, we provide the detailed derivation for diagonalization in the presence of a non-topological defect. We begin with the initial expansion of the scattering process
\begin{equation}\label{eq:fzleft}
\begin{split}
A_{i}(\theta_i) \hat{\mathbb{A}}_{k,\alpha}^{\{\mu\}} &= T_{i \alpha}(\theta_i) \prod_{l=1}^{N} S_{i\mu_l} \hat{\mathbb{A}}_{k,\alpha}^{\{\mu\}} A_{i}(\theta_i) \\
&+ R_{i \alpha}(\theta_i)  \prod_{l=1}^{k} S_{i\mu_l} S_{\mu_{l}i} A_{i}(-\theta_i) \hat{\mathbb{A}}_{k,\alpha}^{\{\mu\}}.
\end{split}
\end{equation}
The periodicity condition imposes the following relation on the operators
\begin{equation}\label{eq:assump1}
A_{i}(\theta) \hat{\mathbb{A}}_{k,\alpha}^{\{\mu\}} = \hat{\mathbb{A}}_{k,\alpha}^{\{\mu\}} A_{i}(\theta) e^{-i L m_{i} \sinh \theta} .
\end{equation}
Alternatively, this property can be expressed for the reflected rapidity component as
\begin{equation}\label{eq:assump2}
A_{i}(-\theta) \hat{\mathbb{A}}_{k,\alpha}^{\{\mu\}} = \hat{\mathbb{A}}_{k,\alpha}^{\{\mu\}} A_{i}(-\theta) e^{i L m_{i} \sinh \theta} .
\end{equation}
Utilizing the periodicity and continuing the expansion of Eq.~\eqref{eq:fzleft}, we obtain
\begin{equation}
\begin{split}
A_{i}(\theta_i) \hat{\mathbb{A}}_{k,\alpha}^{\{\mu\}} &= T_{i \alpha}(\theta_i) \prod_{l=1}^{N} S_{i\mu_l}  A_{i}(\theta_i)  \hat{\mathbb{A}}_{k,\alpha}^{\{\mu\}} e^{i L m_{i} \sinh \theta_i}\\
&+ R_{i \alpha}(\theta_i)  \prod_{l=1}^{k} S_{i\mu_l} S_{\mu_{l}i} \hat{\mathbb{A}}_{k,\alpha}^{\{\mu\}} A_{i}(-\theta_i) e^{i L m_{i} \sinh \theta_i}.
\end{split}
\end{equation}
This step corresponds to the initial scattering event depicted in Fig.~\ref{fig:BAE_non_topo}. The first term represents the particle wrapping around the system and returning to the initial position; we do not need to expand this term further. 

The second term, however, involves $A_{i}(-\theta_i)$, representing a reflection. We proceed by allowing this term to scatter from the right side
\begin{equation}
\begin{split}
&A_{i}(\theta_i) \hat{\mathbb{A}}_{k,\alpha}^{\{\mu\}} = T_{i \alpha}(\theta_i) \prod_{l=1}^{N} S_{i\mu_l} e^{i L m_{i} \sinh \theta_i} A_{i}(\theta_i)  \hat{\mathbb{A}}_{k,\alpha}^{\{\mu\}} + R_{i \alpha}(\theta_i)  \prod_{l=1}^{k} S_{i\mu_l} S_{\mu_{l}i} \\
\times & e^{i L m_{i} \sinh \theta_i}\left( R_{i \alpha}(\theta_i) \prod_{l=k+1}^N S_{\mu_{l}i} S_{i\mu_l}  \hat{\mathbb{A}}_{k,\alpha}^{\{\mu\}} A_{i}(\theta_i) + T_{i \alpha}(\theta_i) \prod_{l=1}^N S_{\mu_{l}i} A_{i}(-\theta_i)  \hat{\mathbb{A}}_{k,\alpha}^{\{\mu\}} \right).
\end{split}
\end{equation}
After simplification, we arrive at the following equation
\begin{equation}
\begin{split}
A_{i}(\theta_i) \hat{\mathbb{A}}_{k,\alpha}^{\{\mu\}} &= T_{i \alpha}(\theta_i) \prod_{l=1}^{N} S_{i\mu_l} e^{i L m_{i} \sinh \theta_i} A_{i}(\theta_i) \hat{\mathbb{A}}_{k,\alpha}^{\{\mu\}} \\
&+ R_{i \alpha}(\theta_i) R_{i \alpha}(\theta_i) \prod_{l=1}^{N} S_{i\mu_l} \prod_{l=1}^{N} S_{\mu_{l}i} e^{2 i L m_{i} \sinh \theta_i} A_{i}(\theta_i)  \hat{\mathbb{A}}_{k,\alpha}^{\{\mu\}} \\
&+ R_{i \alpha}(\theta_i) T_{i \alpha}(\theta_i) \prod_{l=1}^{k} S_{i\mu_l} S_{\mu_{l}i} \prod_{l=1}^N S_{\mu_{l}i}  e^{2 i L m_{i} \sinh \theta_i}  A_{i}(-\theta_i) \hat{\mathbb{A}}_{k,\alpha}^{\{\mu\}}.
\end{split}
\end{equation}
The first and second terms reproduce the contributions for the purely transmissive case and the boundary reflection case, respectively. The primary challenge lies in the third term due to the presence of $A_{i}(-\theta_i)$, which hinders the diagonalization.

The strategy is  to iteratively expand this term until the coefficient of $A_{k,\alpha}^{\mu_1\cdots \mu_{k}} A_{i}(-\theta_i)$ becomes negligible. The useful identity for scattering to the left side is
\begin{equation}\label{eq:fzright}
\hat{\mathbb{A}}_{k,\alpha}^{\{\mu\}} A_{i}(-\theta_i) = R_{i \alpha}(\theta_i) \prod_{l=k+1}^N S_{\mu_{l}i} S_{i\mu_l}  \hat{\mathbb{A}}_{k,\alpha}^{\{\mu\}} A_{i}(\theta_i) + T_{i \alpha}(\theta_i) \prod_{l=1}^N S_{\mu_{l}i} A_{i}(-\theta_i)  \hat{\mathbb{A}}_{k,\alpha}^{\{\mu\}} ,
\end{equation}
Applying this to the third term yields
\begin{equation}
\begin{split}
&R_{i \alpha}(\theta_i) T_{i \alpha}(\theta_i) \prod_{l=1}^{k} S_{i\mu_l} S_{\mu_{l}i} \prod_{l=1}^N S_{\mu_{l}i}  e^{2 i L m_{i} \sinh \theta_i} \hat{\mathbb{A}}_{k,\alpha}^{\{\mu\}} A_{i}(-\theta_i) \\
= &R_{i \alpha}(\theta_i) T_{i \alpha}(\theta_i) R_{i \alpha}(\theta_i) \prod_{l=1}^{N} S_{i\mu_l} S_{\mu_{l}i}^2  e^{3 i L m_{i} \sinh \theta_i} A_{i}(\theta_i) \hat{\mathbb{A}}_{k,\alpha}^{\{\mu\}} \\
+ &R_{i \alpha}(\theta_i) T_{i \alpha}(\theta_i) T_{i \alpha}(\theta_i) \prod_{l=1}^{k} S_{i\mu_l} S_{\mu_{l}i} \prod_{l=1}^N S_{\mu_{l}i}^2  e^{2 i L m_{i} \sinh \theta_i} A_{i}(-\theta_i) \hat{\mathbb{A}}_{k,\alpha}^{\{\mu\}} .
\end{split}
\end{equation}
Iterating this procedure $n$ times leads to the expression
\begin{equation}
\begin{split}
A_{i}(\theta_i) \hat{\mathbb{A}}_{k,\alpha}^{\{\mu\}} &= T_{i \alpha}(\theta_i) \prod_{l=1}^{N} S_{i\mu_l} e^{i L m_{i} \sinh \theta_i} A_{i}(\theta_i) \hat{\mathbb{A}}_{k,\alpha}^{\{\mu\}} \\
&+ R_{i \alpha}(\theta_i) \left(\sum_{k=0}^{n} T_{i \alpha}(\theta_i)^k e^{k i L m_{i} \sinh \theta_i}\prod_{l=1}^{N} S_{\mu_{l}i}^k \right) R_{i \alpha}(\theta_i) \\
& \times \prod_{l=1}^{N} S_{i\mu_l} S_{\mu_{l}i} e^{2 i L m_{i} \sinh \theta_i} A_{i}(\theta_i)  \hat{\mathbb{A}}_{k,\alpha}^{\{\mu\}} \\
&+ R_{i \alpha}(\theta_i) T_{i \alpha}(\theta_i)^{n+1} \prod_{l=1}^{k} S_{i\mu_l} S_{\mu_{l}i} \prod_{l=1}^N S_{\mu_{l}i}^{n+1}  e^{(n+1) i L m_{i} \sinh \theta_i} A_{i}(-\theta_i) \hat{\mathbb{A}}_{k,\alpha}^{\{\mu\}} .
\end{split}
\end{equation}
Taking the limit $n \to \infty$, the third term can be dropped after the usual finite-volume analytic continuation or $i0$ regularization of the Bethe-Yang phases. For real rapidities the bulk $S$-matrix has modulus one, so this convergence should be understood with such a regulator, or equivalently by solving the two-component linear system before taking the real-volume limit. The summation in the second term forms a geometric series. Summing this series, the final result is
\begin{equation}
\begin{split}
A_{i}(\theta_i) \hat{\mathbb{A}}_{k,\alpha}^{\{\mu\}} &= T_{i \alpha}(\theta_i) \prod_{l=1}^{N} S_{i\mu_l} e^{i L m_{i} \sinh \theta_i} A_{i}(\theta_i) \hat{\mathbb{A}}_{k,\alpha}^{\{\mu\}} \\
&+ \frac{R_{i \alpha}(\theta_i) R_{i \alpha}(\theta_i)}{1-T_{i \alpha}(\theta_i) e^{i L m_{i} \sinh \theta_i}\prod_{l=1}^{N} S_{\mu_{l}i} } \prod_{l=1}^{N} S_{i\mu_l} S_{\mu_{l}i} e^{2 i L m_{i} \sinh \theta_i} A_{i}(\theta_i)  \hat{\mathbb{A}}_{k,\alpha}^{\{\mu\}}.
\end{split}
\end{equation}
This allows us to extract a secular equation analogous to the standard BAE
\begin{equation}\label{eq:originbae}
T_{i \alpha}(\theta_i) \prod_{l=1}^{N} S_{i\mu_l} e^{i L m_{i} \sinh \theta_i} + \frac{R_{i \alpha}(\theta_i) R_{i \alpha}(\theta_i)}{1-T_{i \alpha}(\theta_i) e^{i L m_{i} \sinh \theta_i}\prod_{l=1}^{N} S_{\mu_{l}i} } \prod_{l=1}^{N} S_{i\mu_l} S_{\mu_{l}i} e^{2 i L m_{i} \sinh \theta_i} =1.
\end{equation}
This quadratic equation yields two possible solutions
\begin{equation}
e^{i L m_{i} \sinh \theta_i}  \frac{1}{2}\left( T_{i \alpha}(\theta_i) (1+ \prod_{l=1}^{N} S_{\mu_l i}^2) \pm \sqrt{T_{i \alpha}(\theta_i)^2(1 - \prod_{l=1}^{N} S_{\mu_l i}^2)^2 + 4 R_{i \alpha}(\theta_i)^2 \prod_{l=1}^{N} S_{\mu_l i}^2} \right)\prod_{l=1}^{N} S_{i\mu_l} = 1.
\end{equation}
Assuming unitarity and parity symmetry, the term $\prod_{l=1}^{N} S_{\mu_l i}^2$ simplifies to unity. Consequently, the equation reduces to the compact form presented in the main text
\begin{equation}
e^{i L m_{i} \sinh \theta_i} \left( T_{i \alpha}(\theta_i) \pm R_{i \alpha}(\theta_i) \right) \prod_{l=1}^{N} S_{i\mu_l} = 1.
\end{equation}

\bibliographystyle{JHEP}
\bibliography{yunfeng.bib}

\end{document}